\documentclass[usenatbib,usegraphicx,twocolumn]{mn2e}
\usepackage[latin1]{inputenc}
\usepackage{graphicx,amsmath}
\usepackage{times}
\usepackage{amssymb}
\usepackage{natbib}
\usepackage{rotating}
\usepackage{lscape}
\usepackage{multirow}
\usepackage{bbold}
\usepackage{gensymb}

\graphicspath{{images/}}

\title[Bayesian lens parametric model selection]{Bayesian approach to gravitational lens model
  selection: constraining $H_0$ with a selected sample of strong lenses}
\author[I. Balm\`es \&
  P.S. Corasaniti]{I. Balm\`es$^{1}$\thanks{E-mail:Irene.Balmes@obspm.fr} \& P.S. Corasaniti$^{1}$\\
$^{1}$CNRS, Laboratoire Univers et Th\'eories (LUTh), UMR 8102 CNRS, Observatoire de Paris,
Universit\'e Paris Diderot,\\ 5 Place Jules Janssen, 92190 Meudon, France}

\begin{document}

\date{}

\maketitle

\label{firstpage}

\begin{abstract}
Bayesian model selection methods provide a self-consistent probabilistic framework to test the validity of competing
scenarios given a set of data. We present a case study application to strong gravitational lens parametric models. 
Our goal is to select a homogeneous lens subsample suitable for cosmological parameter inference.
To this end we apply a Bayes factor analysis to a synthetic catalog of 500 lenses with
power-law potential and external shear. For simplicity we focus on double-image lenses (the largest fraction of lens in the simulated sample)
and select a subsample for which astrometry and
time-delays provide strong evidence for a simple power-law model description. 
Through a likelihood analysis we recover the input value of the Hubble constant to within $3\sigma$ statistical uncertainty. 
We apply this methodology to a sample of double image lensed quasars. 
In the case of B1600+434, SBS 1520+530 and SDSS J1650+4251 the Bayes' factor analysis favors a simple 
power-law model description with high statistical significance. Assuming a flat $\Lambda$CDM cosmology,
the combined likelihood data analysis of such systems gives the Hubble constant $H_0=76^{+15}_{-5}\, {\rm km\,s^{-1}Mpc^{-1}}$ 
having marginalized over the lens model parameters, the cosmic matter density and consistently propagated the observational 
errors on the angular position of the images. The next generation of cosmic structure surveys will provide larger lens 
datasets and the method described here can be particularly useful to select homogeneous lens subsamples adapted to perform 
unbiased cosmological parameter inference.
\end{abstract}


\section{Introduction}
Strong gravitational lenses are powerful cosmological probes. The
distorted images of lensed background sources carry information on 
the distribution of invisible matter in cosmic
structures. Furthermore, as light-rays from distant variable 
sources propagate along differently distorted paths,
time-delays between luminosity variations of the source images test
the underlying cosmological expansion 
\citep[for exhaustive review see][]{Schneider1991,Petters2001,Sahalectures2006}.

The use of lens time-delays as cosmic standard rulers was 
initially proposed by \citet{Refsdal1964,Refsdal1966}. To
date time-delays have been measured only in 21 lensed quasars
out of the few hundreds known strong lens systems. 
Despite the scarcity of observations, lots of effort 
has been devoted to using these data 
to infer the value of the Hubble constant, $H_0$. This is because 
time-delays, differently from other standard cosmological tests, 
do not rely on local calibration measurements, such as
the distance ladder method \citep[for an alternative method see][]{Chavez2012}. 
In the next decade, additional lens time-delay data will become available thanks to observational
programs such as the COSMOGRAIL\footnote{http://www.cosmograil.org} project collaboration, as well as the ILMT project\footnote{http://www.aeos.ulg.ac.be/LMT/index.php}. 
Furthermore, the next generation of cosmic structure surveys such as the Dark Energy Survey (DES), 
the Large Synoptic Telescope (LSST) or the EUCLID satellite mission will detect large sample of strong lens systems 
for which time-delays can be accurately measured through follow-up observations. Time-delays from such datasets 
will be particularly useful to infer cosmological constraints that are complementary to those obtained 
from other cosmic probes \citep[see e.g.][]{Dobke2009,Coe2009,Oguri2010,Linder2011}.

Time-delay measurements are particularly challenging because they require an intense monitoring
of the lens system. However, the main limitation to fully exploit the cosmological information encoded
in such data arises from the uncertainty on the lens mass
distribution as well as the presence of perturbing masses along
the line-of-sight. Because of this, several analyses have focused on
``golden lenses'' \citep{Press1996,Williams1997}. These are systems for which there is
a sufficient number of observational features (e.g. presence of arcs, multiple point-like images, flux ratios
measurements) such as to constrain the lens potential independently of
time-delays. For example, \citet{Wucknitz2004} have studied the lens system B0218+357 and 
inferred $H_0=78\pm 6\,{\rm km\,s^{-1}Mpc^{-1}}$ for a flat Cold Dark 
Matter model with Cosmological Constant ($\Lambda$CDM) with mean matter density $\Omega_m=0.3$. 
\citet{Suyu2010} have analysed the lens B1608+656 and for the same cosmology they
have obtained $H_0=70.6\pm 3.1\,{\rm km\,s^{-1}Mpc^{-1}}$. 

Alternatively one can infer cosmological parameter constraints using
a statistical sample of lens time-delay measurements 
\citep[see e.g.][]{Saha2006,Oguri2007}. Such an approach still requires 
modeling the gravitational potential of each lens in the sample. Nevertheless,
one can hope that systematics due to individual mass model 
uncertainties are averaged out. In such a case sample selection effects 
can be the main source of error \citep*{Oguri2005,Oguri2006}. 

Modeling the lens mass distribution can be distinguished in parametric
\citep[e.g. see][]{Oguri2002,Keeton2003,Oguri2007} and 
non-parametric \citep[e.g. see][]{Kochanek1991,SahaWilliams1997,Koopmans05,Vegetti2009,Suyu2009}
methods. The latter uses linear inversion algorithms to
constrain the lens potential directly from the intensity images of the
lens system, while the former uses the measured properties of the lens to constrain 
a parametrized form of the lens potential. A third approach by \citet{Alard2007,Alard2008} reconstructs 
the lens potential as a perturbative expansion around the Einstein
ring of the lens system. 

Non-parametric methods have been extensively used in a vast literature. As an example a non-parametric reconstruction
algorithm is implemented in Pixelens\footnote{www.qgd.uzh.ch/projects/pixelens/}\citep{WilliamsSaha2000}, 
a numerical code commonly used for lens studies. Using this code \citet{Saha2006}
have found $H_0=72^{+8}_{-11}\,{\rm km\,s^{-1}Mpc^{-1}}$ from a sample
of $10$ time-delay lenses for a flat
$\Lambda$CDM with $\Omega_m=0.3$. Recently \citet{ParaficzHjorth2010}
have performed the same analysis on an extended sample of $18$ lenses 
and obtained $H_0=66^{+6}_{-4}\,{\rm km\,s^{-1}Mpc^{-1}}$. 
A statistical analysis of a sample of lensed quasars using a parametric approach
has been performed by a number of authors. For instance, assuming an isothermal lens
potential \citet{Giovi2001} have inferred limits on $H_0$ for different cosmological models. 
The analysis by \citet{Oguri2007} has propagated the uncertainties in the lens model parameters
and found $H_0=70\pm 6\,{\rm km\,s^{-1}Mpc^{-1}}$ from a sample of $16$ lenses
for a flat $\Lambda$CDM with $\Omega_m=0.26$.
Whether using non-parametric lens reconstruction algorithms or a
parametric model a key aspect of these analyses is the assessment of what
constitutes a good description of the lens potential given the data. 

Independently of the approach used, 
distinguishing between the potentially infinite number of
possibilities has been mostly based on $\chi^2$-statistics.
However, parameter fitting only establishes how well a model
reproduce the data for a given set of model parameter values. 
Deciding whether one model is preferable over another 
is a question of {\it model selection} rather than quality of parameter 
fit. In other words, which model has a higher probability of being the correct model
description of the observations? Do the data justify a more complex
description of the system (additional parameters)? 

In the Bayesian statistical framework these questions are addressed 
by the analysis of the Bayesian evidence and the evaluation of Bayes
factor \citep[][]{Jeffreys1961,MacKay2003,Gregory2005}. 
 Bayesian model selection methods have already been applied to a
 variety of problems in cosmology and astrophysics \citep[e.g.
  see][]{Jaffe1996,Marshall2003,Saini2004,Bassett2004,Mukherjee2006a,Mukherjee2006b,Trotta2007,Ford2007,CornishLittenberg2007,Gregory2010}. 
In the literature, the analysis of Bayes factors has been applied to non-parametric lens
reconstruction techniques \citep{Suyu2006,Vegetti2009} as well as
selecting complicate parametric lens models of galaxy clusters \citep{Jullo2007}.

Here, we present a case study of the use of Bayes factors to construct 
a homogeneous sample of double lensed quasars to
derive bounds on $H_0$. By homogeneous we mean a sample consisting
of lenses whose data, astrometry and time-delays of the images,
provide evidence for the same lens model description at the same level of 
statistical significance. Differently from previous statistical approaches we
consistently propagate all observational uncertainties on the
posterior probabilities, including errors on the angular position of the images,
as well as marginalizing over nuissance model parameters rather than assuming hard priors.
The possibility of selecting homogeneous lens data to infer cosmological parameter
constraints will become particularly important for future survey programs 
which will detect a large number of strong gravitational lenses.
We argue that the use of Bayes factors provides a self-consistent probabilistic
method to build subsample of data which are not dominated by astrophysical selection effects.

The paper is organized as follows: in Section \ref{lenseq} we review the lens equations and the modeling
of the gravitational lens potential, while in Section \ref{bayesanalysis} we discuss the Bayesian model 
selection. We present the results of the lens model selection analysis on simulated data in Section \ref{testbayes} and on real data in Section \ref{modelselection}.
In Section \ref{H0results} we describe the results of the cosmological parameter inference and present our conclusions
in Section \ref{conclusions}.

\section{Gravitational Lenses}\label{lenseq}
\subsection{Lens equations}
Here, we briefly review the basic equations describing the formation
of images in strong gravitational lenses. We assume that the source is 
lensed by an object that can be treated as a single lens plane. 
 
Let us consider an angular coordinate system centered on a lens 
at redshift $z_\rmn{l}$ and a source with angular position $\bbeta$
at redshift $z_\rmn{s}$. According to Fermat's principle 
light-rays from the source follow paths that extremize the
arrival-time. Let $t(\btheta,\bbeta)$ be the arrival-time of 
rays observed at an angle $\btheta$. We can estimate this function 
by considering the geodesics of photons connecting the source-lens
and the lens-observer planes respectively. Two effects
contribute to the arrival-time, a geometrical term 
which accounts for the different length of the lensed paths 
and a gravitational term due to the Shapiro effect \citep[for a detailed derivation e.g. see][]{Blandford86}.
The geometrical contribution reads as 
\begin{equation}
  t_\rmn{geom}(\btheta,\bbeta) = \frac{1+z_\rmn{l}}{2} \frac{D_\rmn{l} D_\rmn{s}}{c D_\rmn{ls}} (\btheta - \bbeta)^2 ,
\end{equation}
where $c$ is the speed of light, while $D_l$, $D_s$ and $D_{ls}$ are the angular diameter distances between
observer and lens, observer and source, and lens
and source respectively; in the following, we note $\theta$ and $\beta$ the norms of
$\btheta$ and $\bbeta$. The gravitational term is given by
\begin{equation}
  t_{\rmn{grav}}\left(\btheta,\bbeta\right) = - 8 \pi \frac{1+z_l}{c^3} \Psi(\btheta)
\end{equation}
where $\Psi(\btheta)$ is a 2-D gravitational potential generated
 by the projected surface mass density on the lens plane at the
 angular position of the source, $\Sigma(\btheta)$, as given by the Poisson-like equation
\begin{equation}
  \bigtriangledown^2 \Psi(\btheta) = G\Sigma(\btheta),\label{poisson}
\end{equation}
where $G$ is the Newton gravitational constant.
It is useful to introduce the critical density
$\Sigma_\rmn{c}=\frac{c^2}{4\pi G}\frac{D_\rmn{s}}{\bold{D_\rmn{l}}D_\rmn{ls}}$,
the convergence $\kappa=\Sigma/\Sigma_\rmn{c}$ and 
the reduced potential $\psi=2 \Psi / (G \Sigma_\rmn{c})$ to
write the total arrival-time as 
\begin{equation}
  t(\btheta, \bbeta) = \label{time} (1+z_\rmn{l}) \frac{D_\rmn{l} D_\rmn{s}}{c D_\rmn{ls}} \left[ \frac{1}{2} (\btheta - \bbeta)^2 - \psi(\btheta) \right],
\end{equation}
from which we can derive all relevant equations describing the properties of
the images. 

\textbf{Lens Equation}: let us assume the source to be fixed at
the angular location $\bbeta$, the arrival-time only depends on $\btheta$ and images will
form at extrema of the arrival time corresponding to $\bigtriangledown t=0$. Using Eq.~(\ref{time})
we obtain the lens equation
\begin{equation}
 \bbeta = \btheta - \bigtriangledown \psi(\btheta).\label{lensequation}
\end{equation}

\textbf{Magnification}: gravitational lensing deforms the images of a lensed source  
and since the luminosity per surface area is conserved, images can appear magnified or demagnified. 
The amplitude of the effect can be quantified by considering the relation between the
coordinates on the source plane and the lens one. Let
\begin{equation}
\mathcal{A}\equiv\bigtriangledown\bbeta=\mathbb{1}-\bigtriangledown\bigtriangledown\psi,
\end{equation}
be the Jacobian matrix of the angular coordinate transformation, this relation
implies that a solid-angle element $\delta\beta^2$ of the
source is mapped to a solid-angle element of the image
$\delta\theta^2$ by the inverse of the determinant of $\bigtriangledown\bbeta$. 
Hence, the magnification is given by
\begin{equation}
\mu\equiv\frac{\delta\theta^2}{\delta\beta^2}=\frac{1}{{\rm det}\mathcal{A}}.\label{mag}
\end{equation}

\textbf{Time-delay}: let us consider a double-image lensed source characterized by a time-variable luminosity, 
the time-delay in the appearance of the luminosity variation between two images A and B is given by:
\begin{eqnarray}
\Delta t_\rmn{AB}&=&t(\btheta_A,\bbeta)-t(\btheta_B,\bbeta)
=(1+z_\rmn{l}) \frac{D_\rmn{l} D_\rmn{s}}{c D_\rmn{ls}} \times \nonumber\\
&\times&\left[\frac{1}{2} (\btheta_\rmn{A} - \bbeta)^2 - \psi(\btheta_\rmn{A}) - \frac{1}{2} (\btheta_\rmn{B} - \bbeta)^2 + \psi(\btheta_\rmn{B}) \right]\nonumber.\\\label{dt}
\end{eqnarray}

The position of the images, their relative flux ratios and time-delays are the main observable features of a strong lens systems. 

\subsection{Lens models}\label{lensgalmod}
Our goal is to select a homogeneous sample of simple lens systems that can be easily
observed in large sky surveys. For this reason we focus on individual
galaxies which lens distant quasar sources.  
 
\subsubsection{Power-law density profile}
The mass distribution of lens spiral and elliptical galaxies is well approximated by 
power-law density profiles \citep[e.g. see][]{Rusin2003} for which the lens potential assume
the form:
\begin{equation}
\psi(\btheta)=\frac{b^2}{3-n}\left(\frac{\theta}{b}\right)^{3-n},\label{pow}
\end{equation}
where $b$ is a deflection scale. The singular isothermal sphere (SIS) model \citep{Binney1987}
corresponds to $n=2$, for which $b=4\pi D_\rmn{ls}\sigma^2/D_\rmn{s}$, where $\sigma$ is
the velocity dispersion of the galaxy. Measurements of galaxy density profiles indicates 
that the slope parameter $n$ is generally close to the isothermal value, 
though some systems have revealed shallow profiles with $n<1$ \citep[e.g. see][]{Salucci2007}.
By computing the second derivative of Eq.~(\ref{pow}) with respect to $\theta$ we obtain the convergence:
\begin{equation}
\kappa(\btheta)=\frac{2-n}{2}\left(\frac{\theta}{b}\right)^{1-n}.\label{convergence}
\end{equation}
In the case of a double image lens with point-like images located at $\theta_A$ and $\theta_B$, the deflection
scale can be written as:
\begin{equation}
b=\left(\frac{\theta_A+\theta_B}{\theta_A^{2-n}+\theta_B^{2-n}}\right)^{\frac{1}{n-1}},
\end{equation}
where we have used Eq.~(\ref{lensequation}).
For such a system the time-delay between the two images is given by
\citep{Kochanek2002}:
\begin{eqnarray}
\Delta{t}_\rmn{AB}&=&(1+z_\rmn{l}) \frac{D_\rmn{l} D_\rmn{s}}{c D_\rmn{ls}} \times \nonumber\\
&\times&\biggl\{\left(1-\langle\kappa\rangle\right)\left[\frac{1}{2}\left(\theta_B^2-\theta_A^2\right)+\theta_A\theta_B\log\left(\frac{\theta_A}{\theta_B}\right)\right]+\nonumber\\
&-&2\int_{\theta_A}^{\theta_B}\left[\kappa(\theta)-\langle\kappa\rangle\right]\log\left(\frac{\theta}{\theta_B}\right)\theta\,d\theta\biggr\}
\end{eqnarray}
where 
\begin{equation}
\langle\kappa\rangle=\frac{2}{\theta_B^2-\theta_A^2}\int_{\theta_A}^{\theta_B}\kappa(\theta)\,\theta
d\theta,
\end{equation}
is the mean surface mass density in the ring between $\theta_A$ and
$\theta_B$. The above formulae can be rewritten using Eq.~(\ref{convergence}):
\begin{eqnarray}
\Delta{t}_\rmn{AB}&=& (1+z_\rmn{l}) \frac{D_\rmn{l} D_\rmn{s}}{c D_\rmn{ls}} \bigg\{\left[\frac{1}{2}-\langle\kappa\rangle\frac{2-n}{3-n}\right] \left[\theta_\rmn{B}^2-\theta_\rmn{A}^2\right]+ \nonumber \\
 &+& \bigg[b^{n-1} \frac{\theta_\rmn{A}^{3-n}\theta_\rmn{B}^2-\theta_\rmn{B}^{3-n}\theta_\rmn{A}^2}{\theta_\rmn{B}^2-\theta_\rmn{A}^2} 
 - (1 - \langle\kappa\rangle)\theta_\rmn{A}\theta_\rmn{B} \bigg]\ln\frac{\theta_\rmn{A}}{\theta_\rmn{B}}  \bigg\}\nonumber \\ \label{timedelay}
\end{eqnarray}
and
\begin{equation}
\langle\kappa\rangle = b^{n-1}\frac{\theta_\rmn{B}^{3-n}-\theta_\rmn{A}^{3-n}}{\theta_\rmn{B}^2+\theta_\rmn{A}^2},\label{avkappa}
\end{equation}
which explicitely depend on the lens model parameters. 

Similarly we can write the flux-ratios between images A and B in terms 
of their angular positions and the parameters of the power law potential using Eq.~(\ref{mag}). The flux-ratio is given by
\begin{equation}
r_\rmn{AB}\equiv\frac{F_A}{F_B},
\end{equation}
where
\begin{equation}
\frac{1}{F_{i}}=\left|\left[1-\left(\frac{\theta_{i}}{b}\right)^{1-n}\right]\left[1-(2-n)\left(\frac{\theta_{i}}{b}\right)^{1-n}\right] \right|,\label{oneoverflux}
\end{equation}
with $i=A,B$. This accounts only for luminosity differences between images due to lensing magnification. 
>From Eq.~(\ref{oneoverflux}) we can see that flux-ratio measurements can strongly constrain the lens model parameters,
thus in combination with time-delays such data can reduce or break the mass-sheet degeneracy and lead to tighter bounds on cosmological
parameters. 

Several source of systematic errors currently limit the use of flux-ratio measurements. In fact, these 
can be affected by inhomogeneous dust extinction across the angular size of the lens system, 
thus altering the contribution of the lens magnification encoded in the flux-ratio measurements.
In principle dust effects can be taken into account through color analysis. On the other hand, rapidly varying lensing effects can be a more
important source of systematic uncertainty that remains hard to model. As shown by \cite{SchildSmith1991} in the case of the double 
image lens Q0957+561 flux-ratios can vary over time. Fine variations on short time scales can be caused by microlensing of the structure of the lensed quasar, while trends over longer periods are more indicative of the mass spectrum of the lens galaxy. 

Radio measurements of flux ratios are primarely affected 
by milli-lensing events, while in the optical both milli-lensing and micro-lensing causes flux anomalies. 
In the latter case flux-ratio estimates from spectroscopic measurements may alleviate the systematic effect due to micro-lensing
as recently pointed out by \citet{Sluse2012}. 

Henceforth, it is not surprising that flux-ratios can be an effective probe of the lens mass distribution \citep[e.g. see][]{Goicoechea2005}, 
but at the same time if unmodelled, flux anomalies introduce dominant systematic errors in the cosmological parameter inference. 
Overcoming these limitations requires a systematic monitoring of the lens systems phased on the lens time-delay as well as an accurate modeling
of the lens inner structure. Such detailed analyses have yet to be performed and are beyond the scope of 
our work. Because of this cosmological constraints inferred in combination with flux-ratio measurements should be taken with a grain of salt. 
In Section \ref{bayesfactoranalysis} we will show results obtained by combining time-delays with flux ratios only for illustrative 
purposes. 

\subsubsection{External shear}
The presence of mass perturbators outside the lens system can directly alter 
the lens potential and introduce uncertainties in the modeling of the lens potential.
This is especially the case if a single lens galaxy is not isolated, rather is part of a group or a cluster of galaxies 
where an external shear field can deform the monopole potential Eq.~(\ref{pow}). Such deformation can be
modeled by adding a quadrupole term and in polar angular coordinates $\btheta=(\theta,\phi)$ this reads as
\begin{equation}\label{shear}
\psi_{\rmn{shear}}(\btheta)=-\frac{1}{2}\gamma\,\theta^2 \cos{2(\phi-\phi_\gamma)},
\end{equation}
where $\gamma$ is an amplitude parameter which measure the strength of
the external shear and $\phi_\gamma$ its direction. Observational constraints on the
external shear have been mainly inferred from the study of multiple image lenses.
As an example, a number of studies of individual quadruple lens systems have shown that the shear amplitude is rather large, $\gamma\approx 0.1-0.3$
\citep{Fischer1998, Kneib2000}. In contrast, earlier expectations were in the range $\sim 0.02-0.05$ \citep{Keeton1997}. 
Using results from N-body simulations in combination with semi-analytic models of galaxy formation \cite{Holder2003} 
have shown that distribution of values of $\gamma$ is nearly-Gaussian with a peak at $\gamma\approx 0.1$ and 
a rapid decay for $\gamma>0.2$. On the other hand \cite{Wong2011} have performed an accurate study of nine lens systems 
in galaxy groups and clusters which indicates that on average $\gamma=0.08$, with a distribution 
ranging from $0.02$ to $0.17$. 

Accounting for the effect of external shear as described by Eq.~(\ref{shear}) adds two parameters to the lens model. 
Consequently, Eq.~(\ref{timedelay})-(\ref{oneoverflux}) are modified by terms
which explicitely depends on the external shear parameters. Such terms can be computed analytically using
Eq.~(\ref{shear}), their derivation is quite cumbersome and we leave it to Appendix \ref{appendixA}. 
An important aspect of the presence of external shear concerns the geometrical configuration of the images.
In particular the angular separation of the images $\Theta_\rmn{AB}=\theta_\rmn{B}-\theta_\rmn{A}$, the 
asymmetry $R_\rmn{AB}\equiv|(\theta_\rmn{B}-\theta_\rmn{A})/(\theta_\rmn{B}+\theta_\rmn{A})|$, and the
the deviation from colinearity $\epsilon=|\Theta_\rmn{AB}-180|/180$ are indicators of perturbations about a monopole-like
lens potential. As an example, \cite{Sahalectures2006} have shown that the time-delays have little sensitivity
to the quadrupole term if the images are opposite to each other with respect to the lens position. Similarly, 
\cite{Oguri2007} has shown that for image pairs with an asymmetry parameter 
$R_\rmn{AB}>0.2$ the time-delay is not 
significantly influenced by an external shear, a third-order external perturbation or the presence of dark matter subhalos.
In contrast, for such systems the time-delay is more sensitive to the radial slope of the potential 
(which quantifies the deviation from isothermality), which appears to be the case especially 
for image pairs with a wide angular separation. This offer a first empirical criterion to select a homogeneous lens sample.

In addition to external shear, a non-spherical lens mass distribution can lead to
a non-trivial angular dependence of the lens potential that contributes to the time-delay. If unmodelled this
also introduce a systematic error altering the cosmological parameter inference. In the case of lenses
with images lying opposite to each other the contribution from this internal shear appears only as a second order
correction to the time-delay \cite{Kochanek2002}. This suggests that in general the presence of internal shear should not
be neglected.

\section{Bayesian Statistical Analysis}\label{bayesanalysis}
Our goal is to use Bayesian model selection to construct a
statistically homogeneous sample of gravitational lenses
from which to infer cosmological constraints using time-delays.
More specifically, we aim to select lenses for which data provide strong evidence in favor
of a simple power-law model description and exclude with high statistical significance the
presence of external shear which might contribute to sample selection
effects in single lens galaxies. Here we review the basic elements of
Bayesian statistical analysis. 

\subsection{\label{parameter-estimation}Model parameter estimation and error propagation}
Let us consider a set of observations of a double image
lens $\bmath{D}$, consisting of the angular position of the
images $\theta_i$ ($i=A,B$), the time-delay $\Delta{t}_\rmn{AB}$ (and
 eventually flux ratio $r_\rmn{AB}$). We want to constrain a lens model 
$\mathcal{M}$ specified by a set of model parameters
$\bmath{\alpha}$. For the time being let us assume hard
priors on the cosmological parameters which determine 
the cosmic angular distances in Eq.~(\ref{dt}).
The first step of the data model comparison consists of 
inferring the best fit values of the lens model parameters
as well as their uncertainties expressed in terms of ``credible
intervals''. To this end we promote the model parameters to random
variables and use the Bayes theorem to infer their posterior probability distribution of
the parameters given the data and the model $\mathcal{M}$:
\begin{equation}
P(\bmath{\alpha} \mid \bmath{D},\mathcal{M})=\frac{\tilde{\mathcal{L}}(\bmath{D} \mid \bmath{\alpha},\mathcal{M})P(\bmath{\alpha}\mid\mathcal{M})}{P(\bmath{D} \mid \mathcal{M})},
\end{equation}
where $\tilde{\mathcal{L}}(\bmath{D}\mid\bmath{\alpha},\mathcal{M})$ is the
likelihood function, $P(\bmath{\alpha}\mid\mathcal{M})$ is the prior
parameter model probability and $P(\bmath{D}\mid\mathcal{M})$ is the Bayesian ``evidence''.
Notice that the likelihood is not a probability distribution in the
parameters $\bmath{\alpha}$, since it gives the probability of the
data $\bmath{D}$ for a given value of the model parameters. In contrast,
the evidence is the probability of the observed data within the
assumed model $\mathcal{M}$ and appears as an overall normalization
constant of the posterior. Hence,
for parameter estimation purposes the evidence can be neglected since
we are interested in finding the maximum of the posterior and the
dispersion around it. This can be done by computing 
the likelihood function, which for a set of independent data reads as:
\begin{equation}
\tilde{\mathcal{L}}(\bmath{D} \mid \bmath{\alpha},\mathcal{M})\equiv e^{-\frac{\chi^2}{2}}=\exp\left\{-\sum_i\frac{[D_i-D^i_\textrm{th}(\bmath{\alpha})]^2}{2\,\sigma_{i}^2}\right\},\label{likenoerr}
\end{equation}
where $\sigma_i$ are the observational uncertainties and
$D^i_\textrm{th}(\bmath{\alpha})$ are the model predictions. However, notice
that in our case, the position of the images, $\theta_A$ and $\theta_B$, play the
role of independent variables in $\Delta{t}_\rmn{AB}$ and
$r_\rmn{AB}$ respectively. Since, the measured angles are known up to
observational errors, it is important to propagate the angular uncertainties as well. 
This can be consistently done in the Bayesian framework. For a detailed
discussion on the subject we refer to the exhaustive article by D'Agostini \citep{Dagostini}. 

Let assume that the observed image positions,  $\theta_A$ and
$\theta_B$, are determined with Gaussian uncertainties $\sigma_A$ and
$\sigma_B$ respectively. Let us indicate with $\bold{\Theta}_A$ and $\bold{\Theta}_B$ the
true location of the images. Then, we can propagate the uncertainty on
the independent variables by marginalizing the likelihood over the angular error distribution.
Assuming Gaussian errors this gives (for conciseness we drop the dependence on $\mathcal{M}$):
\begin{equation}
\mathcal{L}(\bmath{D} \mid \bmath{\alpha})= \iint
 f(\bold{\Theta}_A,\bold{\Theta}_B)\,\tilde{\mathcal{L}}(\bmath{D} \mid \bold{\Theta}_A,\bold{\Theta}_B,\bmath{\alpha})\,d\bold{\Theta}_A d\bold{\Theta}_B,\label{likeerr}
\end{equation}
with
\begin{equation}
f(\bold{\Theta}_A,\bold{\Theta}_B)=\exp\left\{-\sum\limits_{A,B}\frac{(\theta_i-\bold{\Theta}_i)^2}{2\sigma^2_i}\right\},\label{likegauss}
\end{equation}
where in the integrand we have explicited the dependence of the likelihood $\tilde{\mathcal{L}}$ on
the position of the images. If the angles are precisely measured, 
then Eq.~(\ref{likegauss}) tends to a $\delta$-Dirac and we recover the
standard likelihood expression Eq.~(\ref{likenoerr}). 

\subsection{Bayesian model selection}
Parameter estimation only provides us with information on the
quality of fit of a given model against the data. How do we choose between
competing models characterized by different model parameters?
Preference solely based on the goodness-of-fit as measured by the
$\chi^2$ for the best fitting model parameter values leaves us with partial
information which is not consistently quantifiable in a probabilistic
manner. Even then, how do we discount for the level of
predictiveness of different models? More precisely, how
do we decide whether data justify the choice of a more complex model 
$\mathcal{M}_1$ over one with a
smaller number of free parameters $\mathcal{M}_0$, while accounting for the extended
prior parameter space of $\mathcal{M}_1$? As stressed in \cite{Liddleetal07}, such a
question is related to the predictiveness of models rather than
`simplicity/complexity', since the former is not necessarily related
to the number of free parameters. Parameter estimation does not address these questions, which require a further
step already built in the Bayesian approach. 

Central object of the Bayesian model selection is the evidence,
\begin{equation}
P(\bmath{D} \mid \mathcal{M})=\int d\bmath{\alpha} \,\mathcal{L}(\bmath{D} \mid \bmath{\alpha},\mathcal{M})P(\bmath{\alpha}\mid\mathcal{M}),
\end{equation}
which gives the probability of the data given the model
$\mathcal{M}$. Using the Bayes theorem we then have the
probability of the model $\mathcal{M}$ given the data,
\begin{equation}
P(\mathcal{M} \mid \bmath{D})\propto P(\bmath{D} \mid \mathcal{M})P(\mathcal{M}),
\end{equation}
where $P(\mathcal{M})$ is the prior belief on the model. Thus, given
two competing models, $\mathcal{M}_0$ and $\mathcal{M}_1$, we can base
our preference on the ratio of the model probabilities given the same
set of data
\begin{equation}
\frac{P(\mathcal{M}_0 \mid \bmath{D})}{P(\mathcal{M}_1 \mid \bmath{D})}=\frac{P(\bmath{D} \mid \mathcal{M}_0)}{P(\bmath{D} \mid \mathcal{M}_1)}\frac{P(\mathcal{M}_0)}{P(\mathcal{M}_1)},
\end{equation}
where the ratio 
\begin{equation}
B_{01}=\frac{P(\bmath{D} \mid \mathcal{M}_0)}{P(\bmath{D} \mid \mathcal{M}_1)},
\end{equation}
is the so called ``Bayes factor'' of $\mathcal{M}_0$ to
$\mathcal{M}_1$. Supposing we have the same prior belief on the two
models, then the Bayes factor gives us an estimate of the ratio of the
model probabilities given the data. Since the evidence sets 
up a tension between the quality of fit of a given model and 
its prior predictiveness, then the Bayes factor 
accounts for the different size of the prior
parameter space of the competing models. 

Bayes factors provide us with a self-consistent probabilistic measure
to select a sample of lenses for which the observational data provide
strong evidence in favor of a given model description. For instance, a simple lens model
is the isothermal sphere which has no free parameters, $\mathcal{M}_0=\{n=2\}$. Using the Bayes factor
we can compare it to a more complex description such as the power-law model 
with one free parameter $\mathcal{M}_1=\{n\}$. In such a case we say that 
$\mathcal{M}_0$ is nested in $\mathcal{M}_1$. Similarly, we can compare 
the power-law model with one free parameter $\mathcal{M}_0=\{n\}$ 
to the case in which a shear field is included with two additional free parameters
$\mathcal{M}_1=\{n,\gamma,\phi_\gamma\}$. The additional presence of internal shear can be included in a third nested model $\mathcal{M}_2$
and confronted against $\mathcal{M}_0$ and $\mathcal{M}_1$. For simplicity, we limit our current analysis to the latter models only. 

In order to assess the strength of
evidence in favor of a model $\mathcal{M}_0$, against a more complex one, $\mathcal{M}_1$, 
we use the Jeffreys scale ~\citep{Jeffreys1961} as reported in ~\cite{Trotta2007}. To be as conservative as possible, 
we select lenses for which $\ln B_{01} > 5$ and exclude those which provide strong evidence in favor of external
shear, i.e. $\ln B_{10} > 5$. We also exclude lenses for which we consider the model comparison to be 
inconclusive $-5 <\ln B_{01}< 5$. Notice that $\ln B_{01} = 5$ corresponds to odds of 1 in about 150
in favor of $M_0$. The arrival of new data update these odds, further improving our knowledge of the lens system. 

\section{Testing Bayesian Model Selection}\label{testbayes}
In this section we perform a detailed analysis on a synthetic lens catalog
to quantify the level of bias on the value of $H_0$ inferred from a Bayes factor selected subsample of simulated lenses. To be as conservative as possible
we consider lenses in the presence of external shear. This allows us to test to which extent Bayes factors
can select those systems for which a simple power-law potential is sufficient to describe the simulated data and consequently evaluate the systematic bias 
on the inferred value of $H_0$.

\begin{figure*}
\begin{tabular}{cc}
\includegraphics[width=8cm]{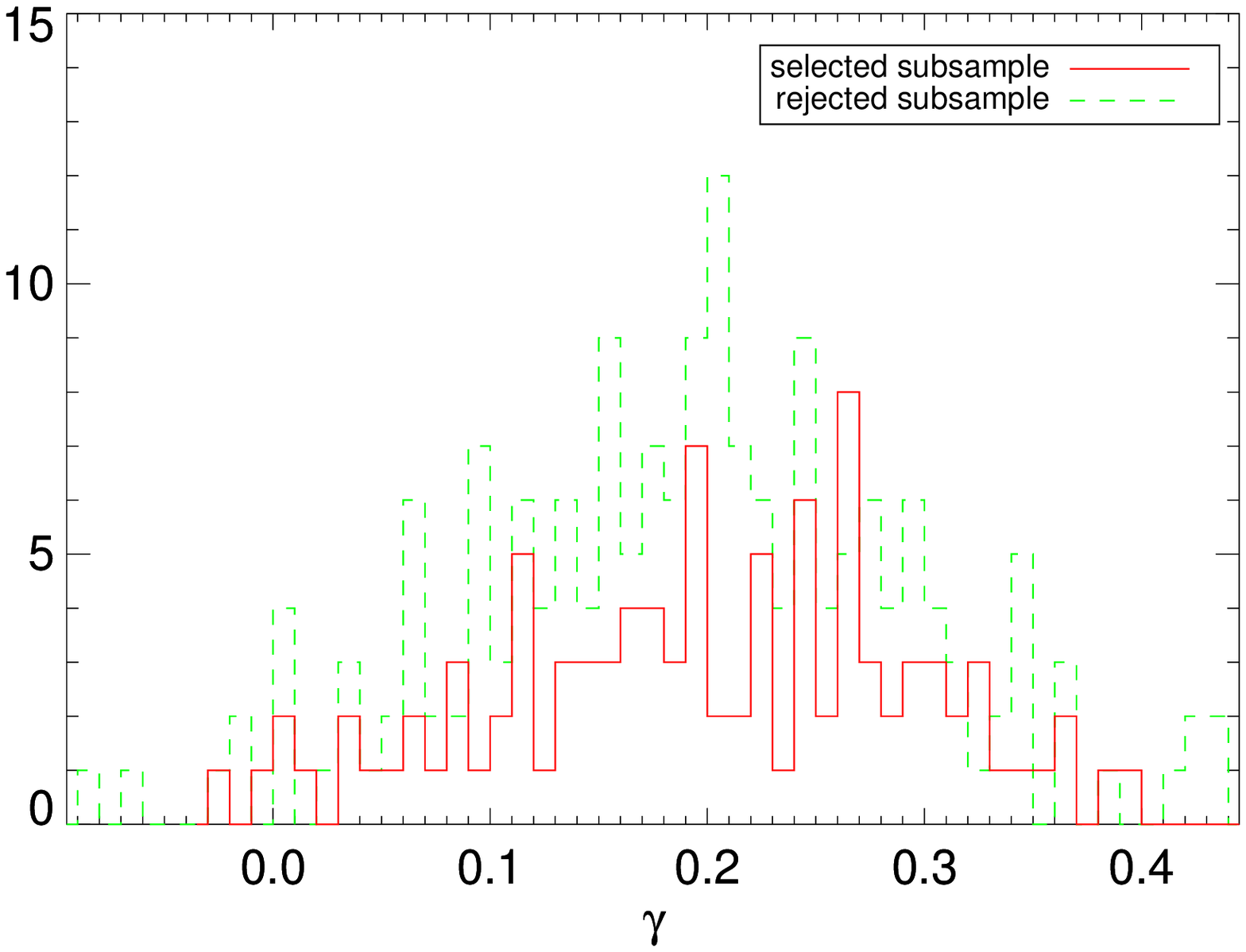}&\includegraphics[width=8cm]{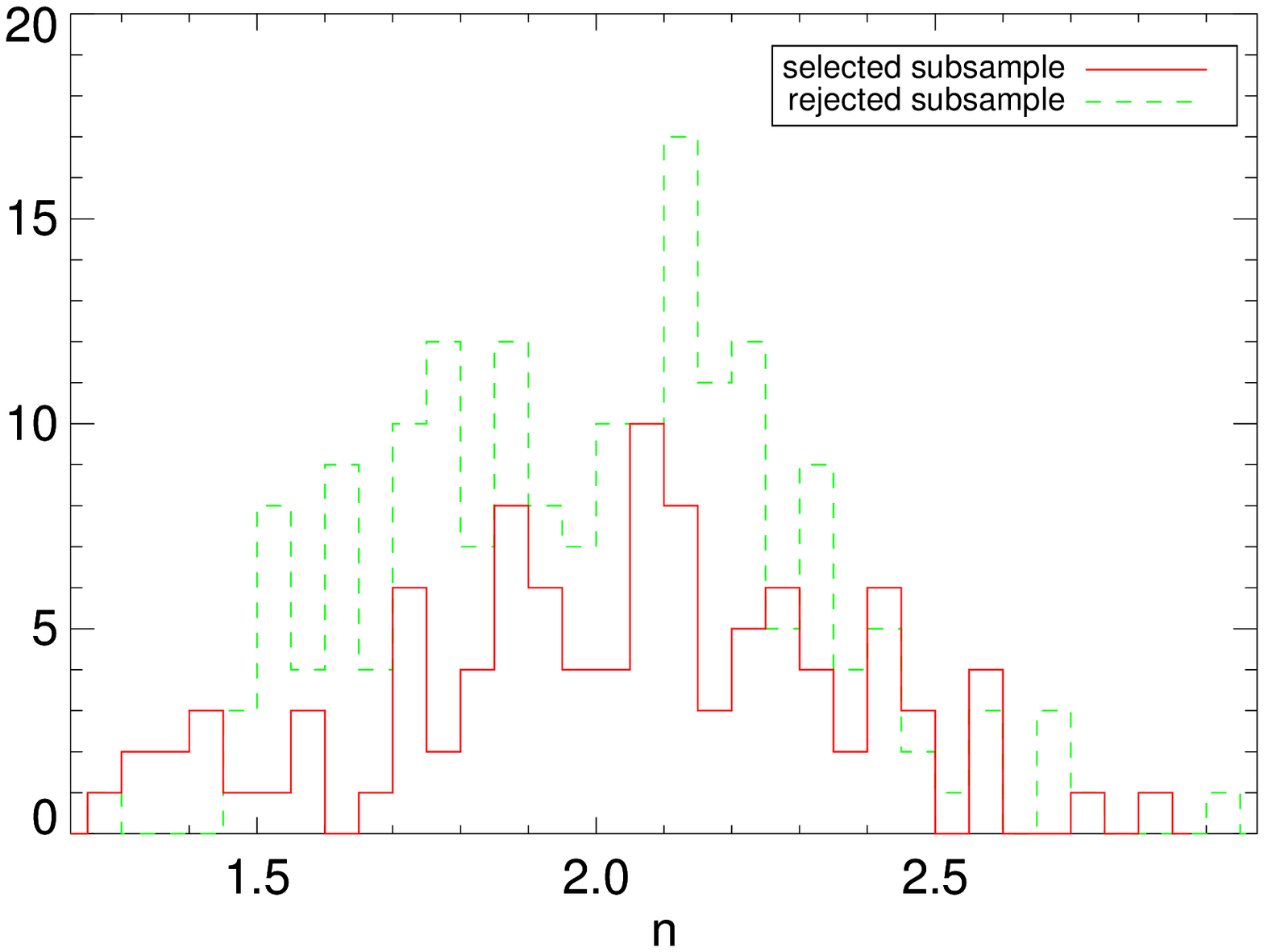}
\end{tabular}
\caption{\label{distribution_gamma} Histogram showing the distribution of simulated values of $\gamma$ (left panel) and $n$ (right panel) for the selected
subsample $\bmath{S_0}$ (solid red line) and the remaining dataset $\bmath{S_1}$ (green dash line).}
\end{figure*}
  
\begin{figure*}
\begin{tabular}{cc}
\includegraphics[width=8cm]{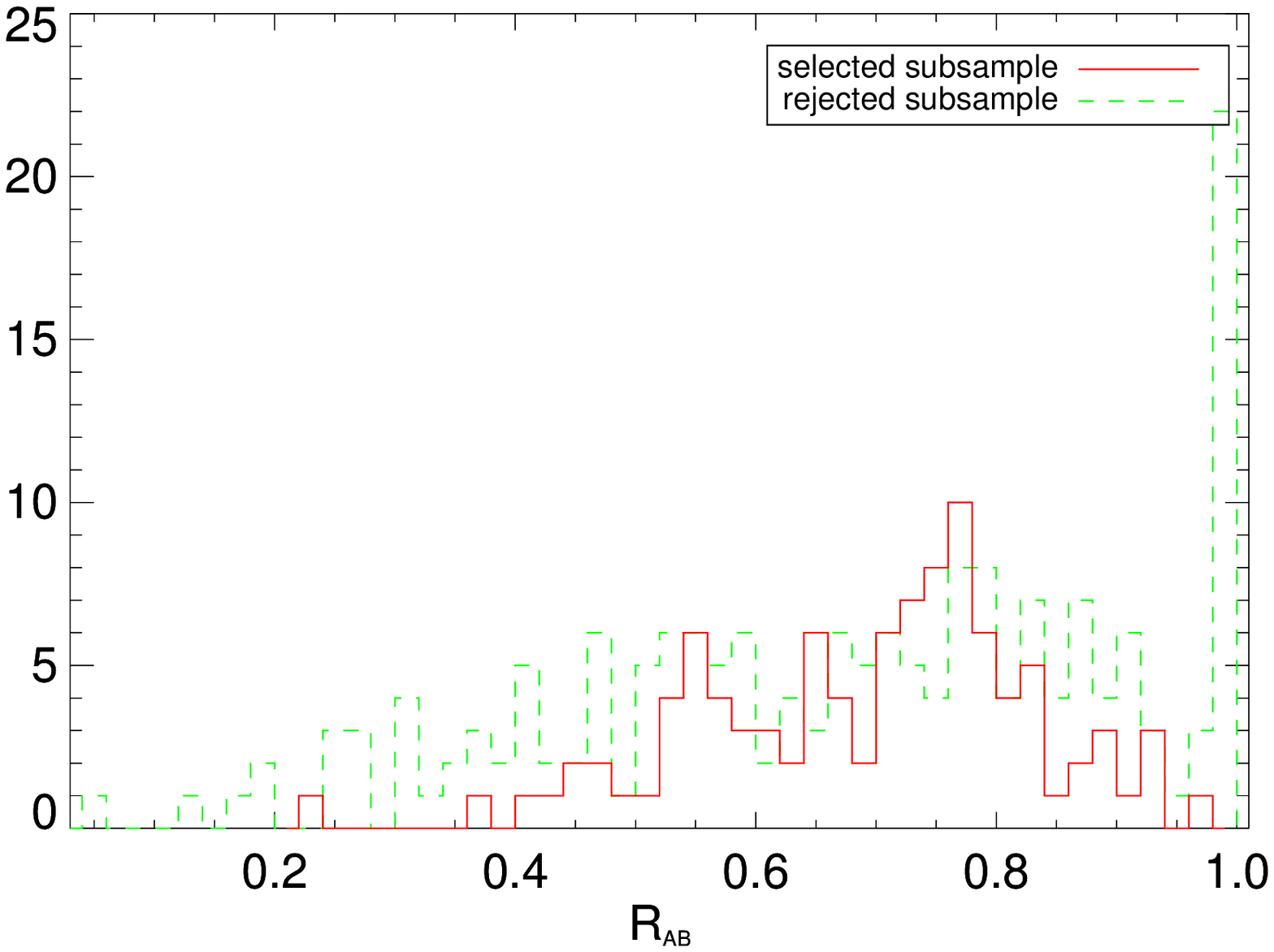}&\includegraphics[width=8cm]{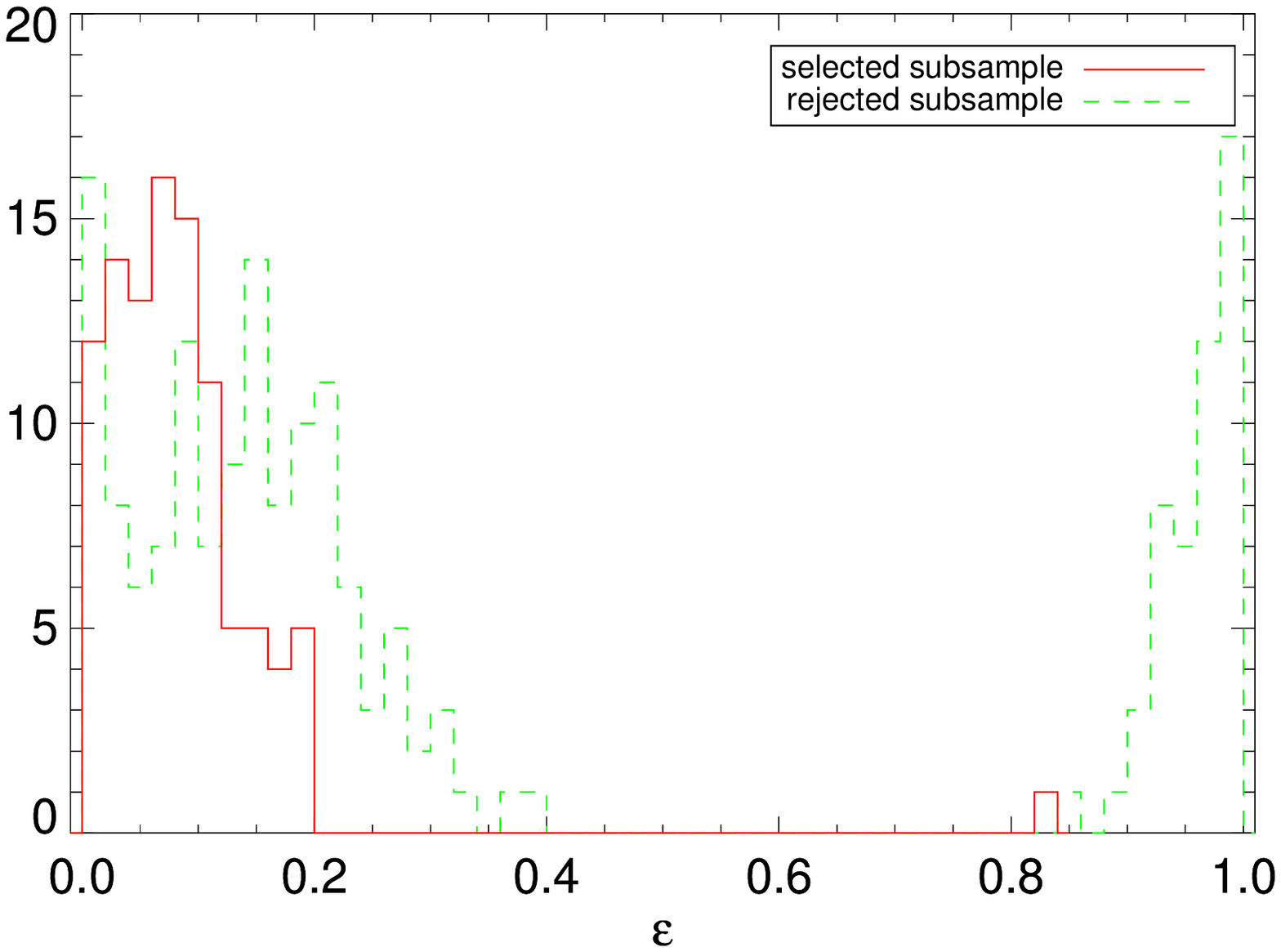}
\end{tabular}
\caption{\label{distribution_ang} Histogram showing the distribution of values of $R_{AB}$ (left panel) and $\epsilon$ (right panel) for the selected
subsample $\bmath{S_0}$ (solid red line) and the rejected dataset $\bmath{S_1}$ (green dash line).}
\end{figure*}

\subsection{Synthetic Lens Catalogue}
We assume a flat $\Lambda$CDM model with $\Omega_m=0.3$ and a reduced Hubble constant $h=0.72$. Using the publicly available software GRAVLENS ~\citep{Keeton2011} we 
generate a sample of 500 gravitational lenses at $z_l=0.5$ with sources at $z_s=1.8$. A more realistic simulation would be to distribute lenses and sources
over a range of redshifts compatible with observations. However, this choice is not relevant to the purpose of this analysis and for simplicity we assume
them to be at the same redshifts. 

The synthetic catalog is built by Monte Carlo generating a distribution of 
values of $n$ and $\gamma$ drawn from Gaussian distributions with $n\sim\mathcal{N}(2,0.3)$ and $\gamma\sim\mathcal{N}(0.2,0.1)$ consistently 
with current observational constraints discussed in Section ~\ref{lensgalmod}. 
The position of sources is drawn from a uniform distribution in a cartesian angular map with coordinates $ -1 < x,\,y < 1$. For simplicity we fix $b=1$. 
For each simulated lens, GRAVLENS computes the number, position and time-delay of the images. Out of 500 simulated lenses only 312 have more than one image. 
In particular, 280 lenses with two or three images (double lenses) and 32 with four or five images (quad lenses). For simplicity we limit our analysis 
to the double lens subsample, since quad lenses require an accurate modeling of the lens angular structure which is essential to explain the formation of four images. 
We assume a $10\%$ error on the time-delay, rather higher than the current mean error, and
neglect angular errors on the image positions.

\subsection{Model Selection and Systematic Bias}
Using the data in the reduced catalog, we compute the likelihood function $\tilde{\mathcal{L}}$ over the prior lens parameter space of 
a simple power-law model $\mathcal{M}_0 = \{n\}$ and one including external shear $\mathcal{M}_1 = \{n,\gamma,\phi_\gamma\}$ respectively.
We assume uniform priors with $n\in(1,3)$ and $\gamma \in (0, 0.2)$. Then, we perform a numerical integration to evaluate the bayesian 
evidence of each model and compute the Bayes factor, assuming a fixed cosmology with $h=0.7$. Given the limited number of parameters we do not 
perform a Monte Carlo sampling of the likelihood, rather we compute it on a fine multi-dimensional grid.

We select all lenses with $\ln B_{01} > 2.5$. This gives us a subsample, $\bmath{S_0}$,
consisting of 111 lenses for which image astrometry and time-delay provide strong evidence in favor of 
a simple power-law model description. We define the remaining lenses as the subsample $\bmath{S_1}$.

In Fig.~\ref{distribution_gamma} is shown the distribution of value of 
$\gamma$ (left panel) and $n$ (right panel) for $\bmath{S_0}$ (solid red line) and the remaining dataset $\bmath{S_1}$ (green dash line).
As we can see the two subsamples cover the same range of values, thus indicating that Bayes factors are not systematically selecting lenses with particular slope profiles or
small external shear amplitude. This might seem at odd with using the selected subsample $\bmath{S_0}$ to perform an unbiased parameter estimation using the simpler model $\mathcal{M}_0$. 
However, as extensively discussed in related literature \citep[e.g. see][]{Mukherjee2006b}, what the Bayes factors do is to put a tension 
between the capacity of the extended model $\mathcal{M}_1$ to better fit the data and its larger prior parameter volume relative to the simpler model. Hence, even though the
selected sample of lenses has non-vanishing shear, the data (image astrometry and time-delay) do not justify the more complex model because 
the gain in fitting with external shear parameters is minimal compared to the size of the enlarged prior parameter space. 
The Bayes factors simply tell us that the simpler model $\mathcal{M}_0$ has to be prefered as it does a better job at describing the data
with a smaller prior parameter space. By construction all simulated lenses have external shear and not surprising more than half of them are indeed discarded 
by the Bayes factor either as inconclusive or as favoring model $\mathcal{M}_1$. 

In Fig.~\ref{distribution_ang} we plot the histogram of the values of $R_{AB}$ and $\epsilon$ for $\bmath{S_0}$ and $\bmath{S_1}$ respectively. We may
notice that in $\bmath{S_0}$ the distribution of values of $R_{AB}$ has a smaller scatter than for $\bmath{S_1}$ with $R_{AB}>0.35$. Similarly the distribution of
values of $\epsilon$ for the subsample $\bmath{S_0}$ is narrower than $\bmath{S_1}$ and limited to values with $\epsilon<0.2$. 
This clearly indicate that the Bayes factor analysis has selected lenses for which the angular configuration of the images is characteristics of double lens system whose 
astrometry and time-delays are less sensitive to external shear effect \citep[e.g. see][]{Oguri2007}, hence successfully described by a simple model. 
Thus, the Bayes factors automatically performs an empirical model selection based on the structural properties of the lens systems. 

\begin{figure}
\includegraphics[width=8cm]{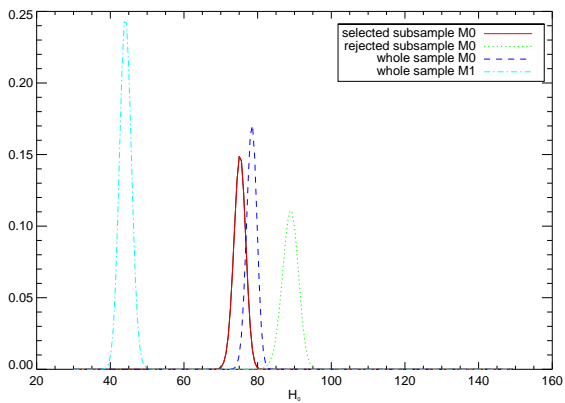}
\caption{\label{simu-likelihoods} Marginalized likelihoods on $H_0$ from the analysis of $\bmath{S_0}$ (solid red line), $\bmath{S_1}$ (green dot line) and 
the entire sample $\bmath{S_0}+\bmath{S_1}$ (dark blue dash line), all with model $\mathcal{M}_0$. 
The dash-dotted light-blue line is obtained by analysing the entire sample with model $\mathcal{M}_1$. This model is underconstrained by time-delay and astrometry data 
only, hence it is not surprising that due to the unbounded parameter degeneracies the inferred value of $H_0$ strongly differs from the fiducial one.}
\end{figure}

We are now in a position to quantify the bias on the inferred value of $H_0$ from the selected lens subsample. Assuming model $\mathcal{M}_0$ we run a likelihood analysis
on $\bmath{S_0}$, $\bmath{S_1}$ and their combination $\bmath{S_0}+\bmath{S_1}$. The marginalized likelihoods on $H_0$ are shown in Fig.~\ref{simu-likelihoods}. 
>From $\bmath{S_0}$ we find $H_0 = 75.94^{+1.5}_{-1.4}\,{\rm km\,s^{-1}Mpc^{-1}}$, quite remarkably this recovers the fiducial value $H_0=72\,{\rm km\,s^{-1}Mpc^{-1}}$ 
to within $3\sigma$ statistical uncertainty of the best fit value, thus indicating a systematic bias of $\sim 5\%$. 
In contrast, the analysis of $\bmath{S_1}$ gives a highly biased value with $H_0 = 90.45^{+2.4}_{-2.1}\,{\rm km\,s^{-1}Mpc^{-1}}$, while
the combined analysis gives $H_0 = 78.39\pm 1.5\,{\rm km\,s^{-1}Mpc^{-1}}$. In principle, the residual $\sim 5\%$ systematic bias in
$\bmath{S_0}$ can be reduced or possible removed using additional lens measurements. Here, we have limited to use time-delay and astrometry data only to 
perform a proof of concept assuming simple lens toy models. However, in a real setup, the availability of the stellar kinematic measurements of the 
lens galaxy as well as high resolution imaging of the Einstein ring may provide additional constraints on the lens mass distribution, thus reducing the internal lens model
parameter degeneracy, which can lead to a more accurate lens model selection and consequently less biased results. 
In fact, the former provides an independent estimate of the lens mass at a radius different from that of the Einstein ring \citep{TreuKoopmansetal}, 
while the thickness of the Einstein ring is sensitive probe of the density profiles \citep[e.g. see][]{Suyu2009}. 
This is to say that the residual systematic bias we have found from the analysis of the simulated lenses
is not inevitable in strong lens cosmography.

One may wonder whether analysing the entire dataset using model $\mathcal{M}_1$ may lead to less biased results given that the data have been generated using such a model. However,
we should remind that model $\mathcal{M}_1$ is underconstrained using double lenses. This can introduce very large parameter degeneracies and eventually 
strongly bias the cosmological parameter inference. To this purpose we have run a likelihood analysis of the whole sample $\bmath{S_0}+\bmath{S_1}$ assuming model $\mathcal{M}_1$.
The marginalized likelihood on $H_0$ is plotted in Fig.~\ref{simu-likelihoods} (dash-dotted light-blue line) from which we obtain $H_0 = 44\pm1.5\,{\rm km\,s^{-1}Mpc^{-1}}$. 
This shows how subtle the parameter inference can be. In a real dataset we will not know a priori whether external shear is indeed present or not. 
Thus, given a set of data which we aim to use to extract cosmological parameter information we are better guided by Bayesian model selection than blindly 
adding parameters to model the complexity behind the data. However, as already stressed above we should not dispair since additional constraints from independent mass proxies
may help reduce or break internal model parameter degeneracies and consequently the robustness of the lens model selection.

\begin{table*}
 \begin{minipage}{175mm}
  \centering
  \caption{\label{data} Double-image lenses. $z_l$ and $z_s$ are the lens and source redshift respectively, $\theta_\rmn{i}$ are the angular position of the images 
relative to the lens position in arcsec, $\Delta t$ is the time-delay (in days) and $F_\rmn{i}/F_\rmn{j}$ the flux-ratios. Observational uncertainties are $1 \, \sigma$ errors. We also quote the derived values of the asymmetry $R_\mathrm{AB}$, angular separation $\Theta_\mathrm{AB}$ and deviation from colinearity $\epsilon$.}
   \begin{tabular}{@{}lcccccccccc}
    \hline
    Lens & pair (i,j) & $z_\rmn{l}$ & $z_\rmn{s}$ & $\theta_\rmn{i}$ ('')   & $\theta_\rmn{j}$ ('')   & $\Delta t=t_\rmn{i}-t_\rmn{j}$ & $\frac{F_\rmn{i}}{F_\rmn{j}}$ & $R_{AB}$ & $\Theta_{AB}$ & $\epsilon$\\ \hline
    
     B0218+357\footnote{Image positions from \citet{Lehar00} computed relative to the lens center as measured by \citet{York2005}; lens and source redshifts from \citet{Browne1993} and \citet{Cohen2003} respectively, 
time-delay from \citet{Biggs1999} and flux ratio from \citet*{Wucknitz2004}.} &  BA  & $0.685$ & $0.944$ & $0.057 \pm 0.004$ & $0.280 \pm 0.008$ & $+10.5 \pm 0.2$ & $0.26 \pm 0.005$ & 0.66 & 205.6 & 0.14\\ 
    
    B1600+434\footnote{Image positions and redshifts from \citet*{Koopmans98}, time-delay from \citet{Burud00}, flux-ratio from \citet{Dai05}.}   &  AB  & $0.414$ & $1.589$ & $1.14 \pm 0.075$   & $0.25 \pm 0.074$   & $-51.0 \pm 2.0$ & $1.75 \pm 0.34$ & 0.45 & 200.2 & 0.11\\ 
    
    FBQ 0951+2635\footnote{Image positions and redshifts as listed in \citet{CASTLES}, time-delay from \citet{Jakobsson05}, flux-ratio from \citet{Shalyapin09}.}  &  AB  & $0.260$ & $1.246$ & $0.886 \pm 0.004$ & $0.228 \pm 0.008$ & $-16.0 \pm 2.0$ & $3.15 \pm 0.05$ & 0.59 & 201.2 & 0.12\\ 
     
    HE 1104--1805\footnote{Image positions and redshifts from \citet{Lehar00}, time-delay and flux-ratio from \citet{Poindexter07}.}   &  AB  & $0.729$ & $2.319$ & $1.099 \pm 0.004$ & $2.095 \pm 0.008$ & $+152.2 \pm 3.0$ & $2.84 \pm 0.06$ & 0.32 & 175.9 & 0.02\\ 
    
    HE 2149--2745\footnote{Image positions and redshifts as listed in \citet{CASTLES}, time-delay and flux-ratio from \citet{Burud02a}.}   &  AB  & $0.603$ & $2.033$ & $1.354 \pm 0.008$ & $0.344 \pm 0.012$ & $-103.0 \pm 12.0$ & $4.0 \pm 0.5$ & 0.59 & 178.9 & 0.006\\ 
     
    PKS 1830--211\footnote{Image positions and redshifts from \citet{Meylan2005}, time-delay from \citet{Lovell1998}, flux-ratio from \citet{Courbin98}.}   &  AB  & $0.89$  & $2.507$ & $0.67 \pm 0.08$   & $0.32 \pm 0.08$   & $-26 \pm 5$ & $1.03 \pm 0.02$ & 0.36 & 199.5 & 0.11\\ 

    Q0142-100\footnote{Image positions and redshifts as listed in \citet{CASTLES}, time-delay and flux-ratio from \citet{Koptelova2012}.}   &  AB  & $0.49$ & $2.719$ & $1.855 \pm 0.002$ & $0.383 \pm 0.005$ & $-89 \pm 11$ & $6.3 \pm 0.1$ & 0.66 & 167.8 & 0.07 \\ 
     
    Q0957+561\footnote{Image positions and redshifts as listed in \citet{CASTLES}, time-delay from \citet{Colley03}, flux-ratio from \citet{Haarsma99}.}      &  AB  & $0.36$  & $1.413$ & $5.220 \pm 0.006$ & $1.036 \pm 0.11$  & $-417.09 \pm 0.07$ & $1.35 \pm 0.04$ & 0.67 & 154.5 & 0.14\\ 
    
    SBS 0909+532\footnote{Image positions and redshifts as listed in \citet{CASTLES}, time-delay from \citet{Ullan06}, flux-ratio from \citet{Dai09}.} &  AB  & $0.830$ & $1.377$ & $0.415 \pm 0.126$ & $0.756 \pm 0.152$ & $+45.0 \pm 5.5$ & $0.32 \pm 0.03$ & 0.29 & 220.3 & 0.22\\
      
    SBS 1520+530\footnote{Image positions and redshifts as listed in \citet{CASTLES}, time-delay from \citet{Burud02b}, flux-ratio from \citet{Auger08}.}   &  AB  & $0.717$ & $1.855$ & $1.207 \pm 0.004$ & $0.386 \pm 0.008$ & $-130.0 \pm 3.0$ & $2.9 \pm 0.4$ & 0.52 & 202.6 & 0.13\\ 
    
    SDSS J1206+4332\footnote{Image positions and redshifts as listed in \citet{Ogurietal2005}, time-delay and flux-ratio from \citet*{Paraficz09}.} & AB & $0.748$ & $1.789$ & $1.870 \pm 0.088$ & $1.278 \pm 0.097$ & $-116 \pm 5$ & $0.74 \pm 0.2$ & 0.19 & 132.9 & 0.26\\
    
    SDSS J1650+4251\footnote{Image positions and redshifts as listed in \citet{CASTLES}, time-delay and flux-ratio from \citet{Vuissoz07}.} &  AB  & $0.577$ & $1.547$ & $0.872 \pm 0.027$ & $0.357 \pm 0.042$ & $-49.5 \pm 1.9$ & $6.2 \pm 0.31$ & 0.42 & 145.8 & 0.19\\ \hline
  \end{tabular}
 \end{minipage}
\end{table*}
 
\section{Application to real lens data}\label{modelselection}
\subsection{Data sample}\label{datasample}
Gravitational time-delays have been measured in 21 strong lens systems. Out of this sample we only consider double 
image lenses in which a far distant quasar is lensed into two images by a foreground galaxy. This reduces our initial dataset to 12 lenses, 
whose characteristics are quoted in Table \ref{data}. The object responsible for the lensing has been
unambigously detected in all listed lenses. However, current observations do not provide clear indications whether the lensing galaxies
are part of a group/cluster or whether perturbators are present along the line of sight. Thus, the presence of 
external shear cannot be a priori excluded, potentially leading to bias selection effects.

For each lens in the sample we compute the 
asymmetry $R_\rmn{AB}$, the angular separation $\Theta_\rmn{AB}$ and the deviation from
colinearity $\epsilon$ as indicators of perturbations about a monopole-like
lens potential. These are also given in Table~\ref{data}, while in Table~\ref{astrometry}
in Appendix~\ref{appastrometry} we report the astrometry.

\subsection{Model parameter priors}
We want to test whether the data listed in Table \ref{data} provide strong evidence for or against a simple 
power-law model, $\mathcal{M}_0=\{n\}$, or one which includes external shear, $\mathcal{M}_1=\{n,\gamma,\phi_\gamma\}$.
Uninformative data cause the Bayes factors to depend on the size of the assumed prior parameter
space. To test for such dependence we distinguish two different sets of priors for each lens model parameter. 

To be as conservative as possible we assume a uniform ``large prior'' on the radial slope of the potential, $n\in(0,3)$ and
a uniform ``small prior'' corresponding to $n\in(1,3)$. The former includes shallow galaxy profiles ($n<1$) such
as those found in e.g. \cite{Salucci2007}, while the latter is limited to the cuspy profiles, including the isothermal
sphere, which are typical of Cold Dark Matter halos. 
Similarly, we assume flat priors on $\gamma$ and consider two prior intervals, $\gamma\in(0,0.1)$ and
$\gamma\in(0,0.2)$, while the external shear orientation angle is assumed $\phi_\gamma\in(0,\pi)$.

In the evaluation of the Bayes factors we consider a flat concordance $\Lambda$CDM model and assume for simplicity
a value compatible with WMAP-7yr \citep{WMAP7} on the matter density, $\Omega_m=0.27$. In order to test the dependence of the Bayes factors on the value of the Hubble constant
we assume three prior values: $H_0=65,70.2$ and $75\,{\rm km\,s^{-1}Mpc^{-1}}$ respectively. 
  
\begin{figure*}
\begin{tabular}{cc}
\includegraphics[width=11cm]{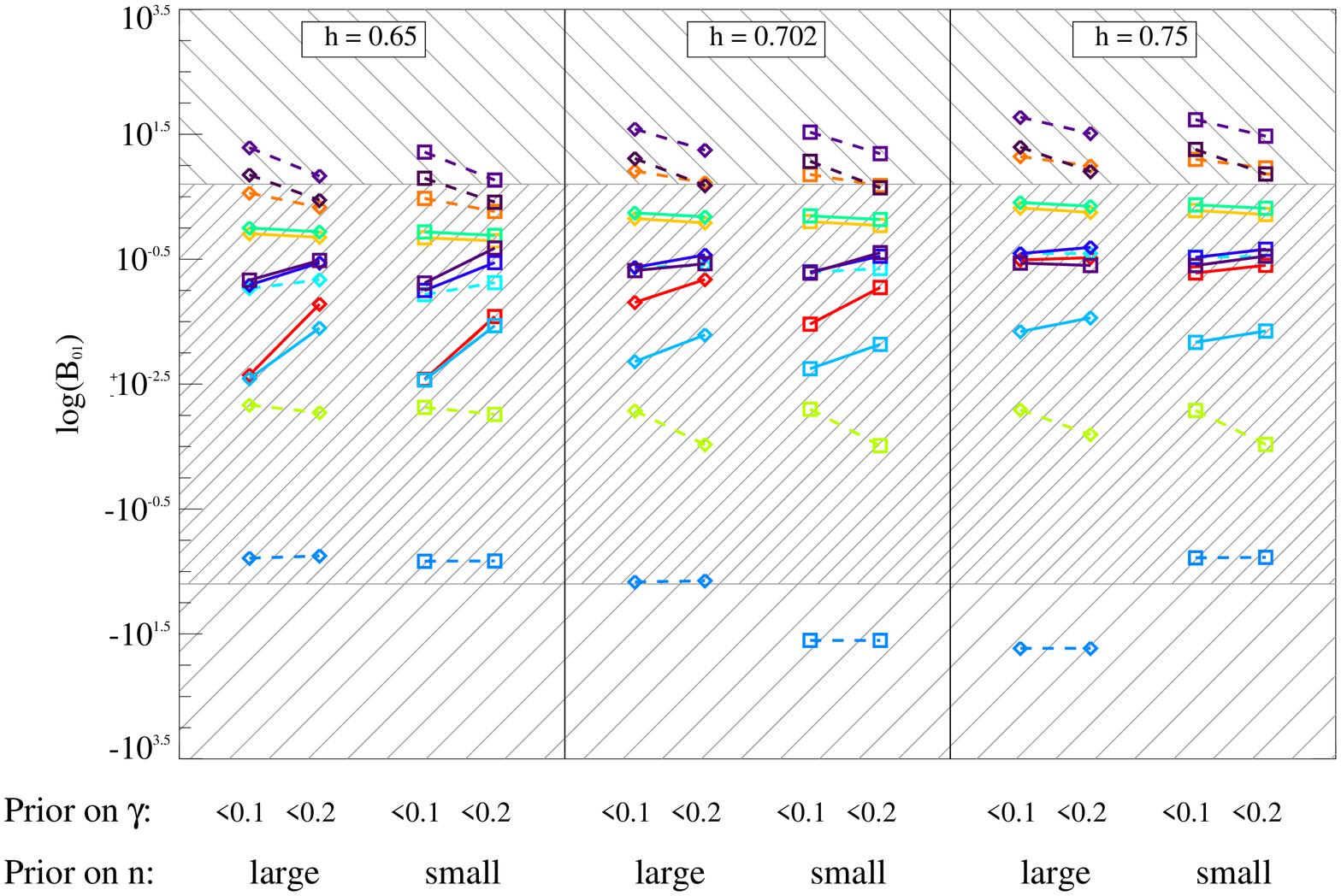}&\includegraphics[width=5cm]{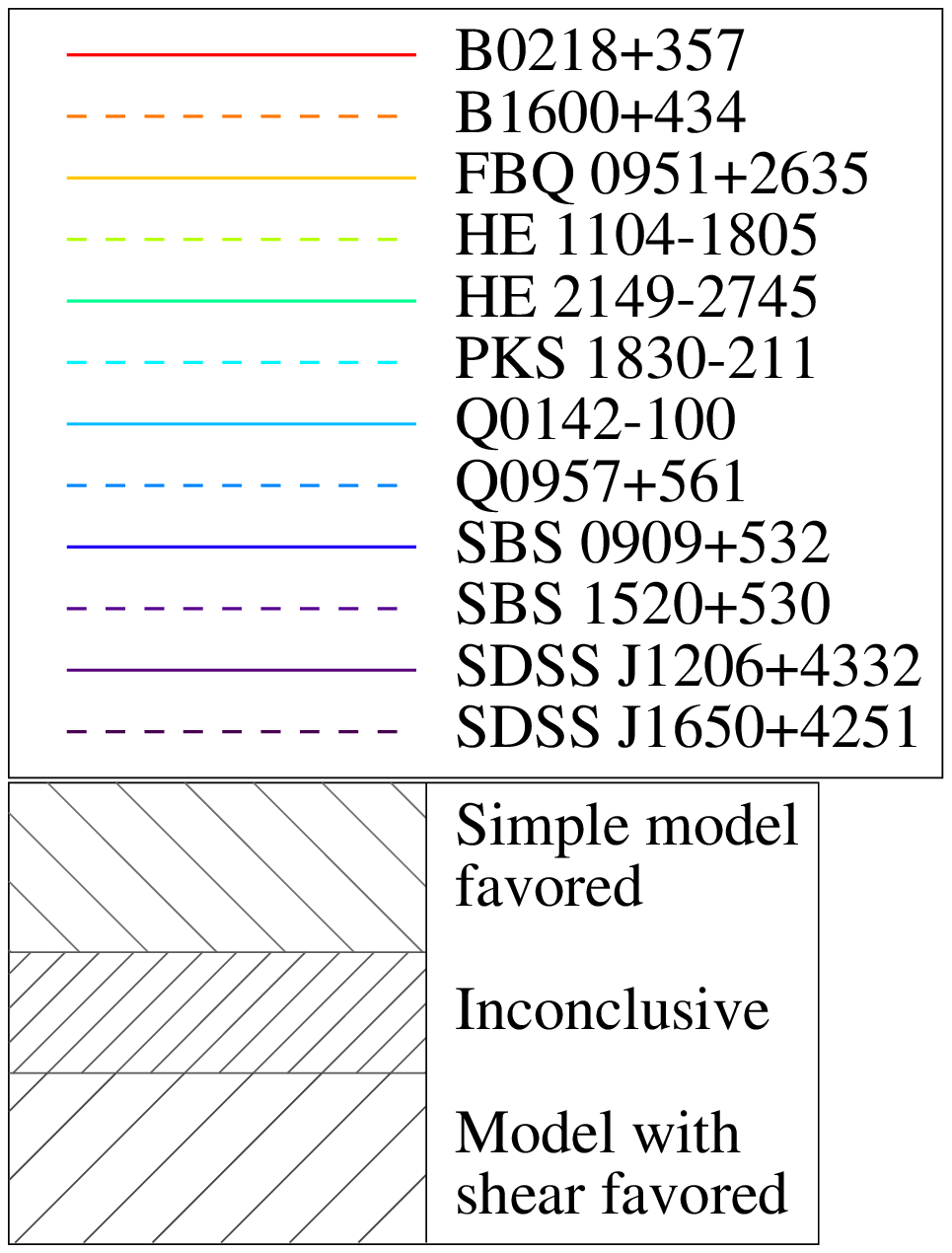}
\end{tabular}
\caption{\label{evidence_dt}Bayes factors for the sample of lenses listed in Table~\ref{data} and obtained from the analysis of the image positions and time-delays for different lens model parameter priors and assuming a hard prior on the reduced Hubble parameter with $h=0.65$ (left panel),
$h=0.702$ (middle panel) and $h=0.75$ (right panel).}
\begin{tabular}{cc}
\includegraphics[width=11cm]{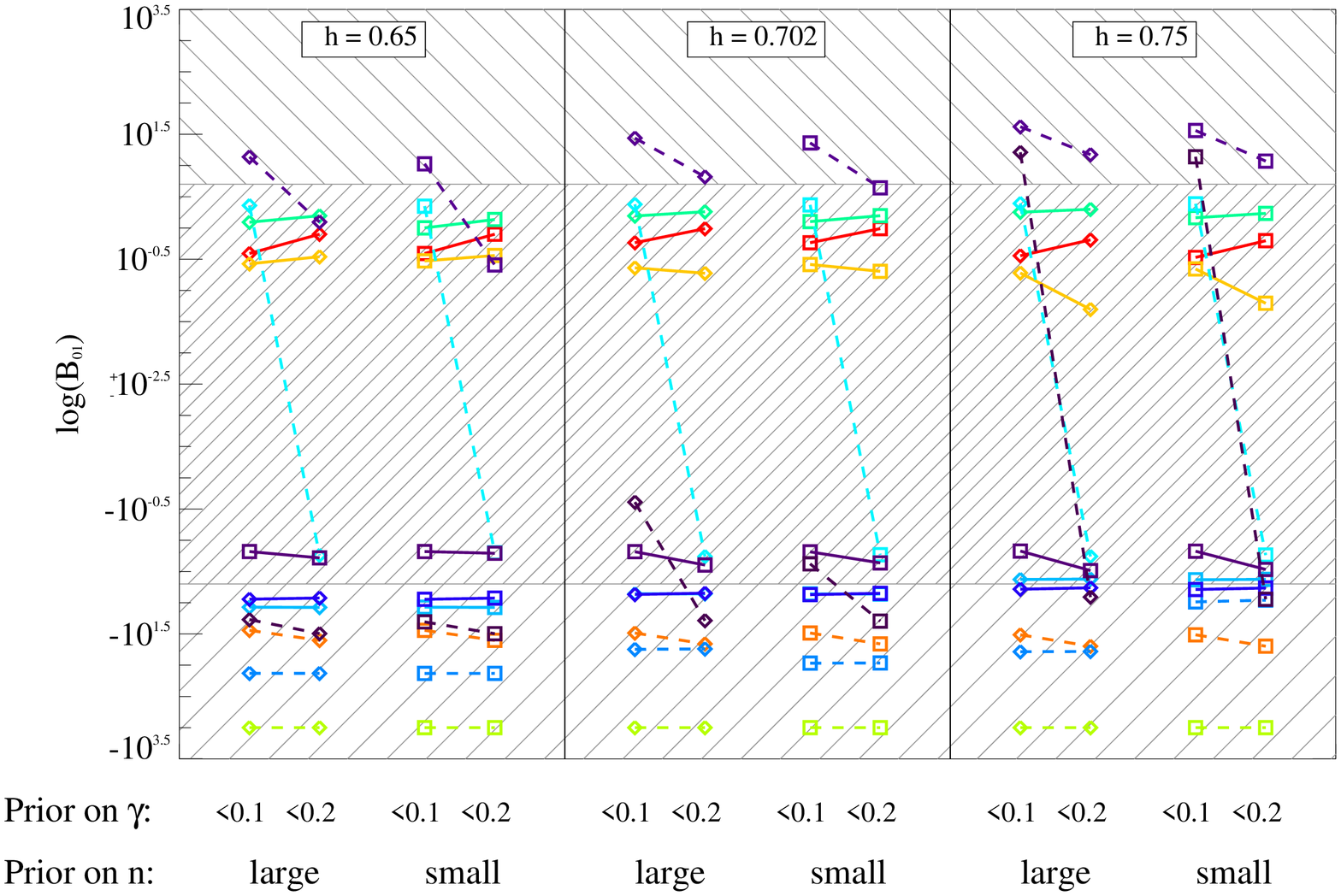}&\includegraphics[width=5cm]{fig6bis_col.eps}
\end{tabular}
\caption{\label{evidence_dtfr} As in Figure \ref{evidence_dt} including information from flux-ratio measurements.}
\end{figure*}

\subsection{Bayes factor analysis}\label{bayesfactoranalysis}
The results of the numerical computation of the Bayes factors obtained from the analysis of time-delays and image astrometry are summarized in Fig.~\ref{evidence_dt}, while in Fig.~\ref{evidence_dtfr}
we show the results obtained including flux-ratio measurements. For each lens in the 
sample we mark the value of $\log_{10}{B_{01}}$ under
different lens model parameter priors and for values of the Hubble constant $H_0=65$ (left panel), $70.2$ (middle panel) and 
$75\,{\rm km\,s^{-1} Mpc^{-1}}$ (right panel) respectively. The hatched areas correspond to values of the 
Bayes factors for which the data provide strong evidence in favor of the simple power-law description ($\ln B_{01}>5$), 
the presence of external shear ($\ln B_{10}>5$). We assume the Bayes model comparison to be inconclusive in the region $-5<\ln B_{01}<5$ which is a very conservative cut.

Let us focus on Fig.~\ref{evidence_dt}. First we can see that Bayes factors 
show no variation for the large and small
priors on $n$, except for Q0957+561. On the other hand, most lenses exhibit prior dependence on $\gamma$, which can vary in strenght with the cosmology. FBQ~0951+2635, HE~2149-2745 and Q0957+561 are the only ones for which the prior on $\gamma$ seems to have no influence whatsoever. We may also notice a weak dependence of the Bayes factors on the value of $h$. In particular, B1600+434 and 
SDSS J1650+4251 have Bayes factor which are clearly above our conservative cut for $h=0.702$ and $h=0.75$, while these shift in the upper bound on the inconclusive interval for $h = 0.65$. 
However, since the value of $B_{01}$ is still rather high, these lenses would pass a slightly less conservative cut even in the worst case scenario of $h=0.65$.

All lenses except HE 1104-1805 and Q0957+561
have Bayes factors favoring $\mathcal{M}_0$, though they may lie in the inconclusive area. 
Q0957+561 is the only lens clearly favoring model $\mathcal{M}_1$ in some case. One final remark concerns SBS~1520+530. 
Our analysis indicates that the image position and the time-delay of this lens strongly favor
a simple power-law lens model without external shear. This is not in contrast with observations of the lens properties by \cite{Auger08} who have shown
the presence of galaxy groups at nearby redshifts on the lens line of sight. It is plausible that mass potential is centered on the lens
(as suggested by the galaxy-groups position relative to the lens) thus implying a negligible quadrupole contribution. Hence, as discussed 
in section ~\ref{testbayes} the extra parameters from modeling the external shear may provide a minimal gain in fitting the data and in such a case
the Bayes factor favours the simpler model description. Furthermore, \citet{Auger08} find that the system
can be well fitted by a nearly isothermal density profile. 

In Fig.~\ref{evidence_dtfr} we plot the Bayes factors obtained from time-delays and flux-ratio measurements. As we can see including flux-ratios
shifts the Bayes factors of several lens systems in the range which strongly favor the presence of external shear. However, it is possible that the extra 
parameters of model $\mathcal{M}_1$ fit unmodeled systematics affecting the flux-ratios rather than the effects of an external shear field. As already mentioned in 
Section \ref{lensgalmod}, current observations already indicate that dust extinction as well as microlensing and substructure alter the flux-ratios. Since
we are far from being able to quantitatively account for such contaminations, it is 
premature to infer any conclusion based on these data. 

In the light of the analysis of Bayes factors inferred from position of the lens images and time-delays 
we select B1600+434, SBS~1520+530 and SDSS~J1650+4251 out of the initial lens sample. 
These are the only lenses which pass our selection criterion independently of the model parameter priors
with the power-law model strongly favored over that including external shear with odds of more than $150$ to $1$. 
In the next Section we will present the constraints on $H_0$ obtained from the likelihood analysis of 
these lenses.

\section{Cosmological Parameter Inference}\label{H0results}
\begin{figure}
\includegraphics[width=8cm]{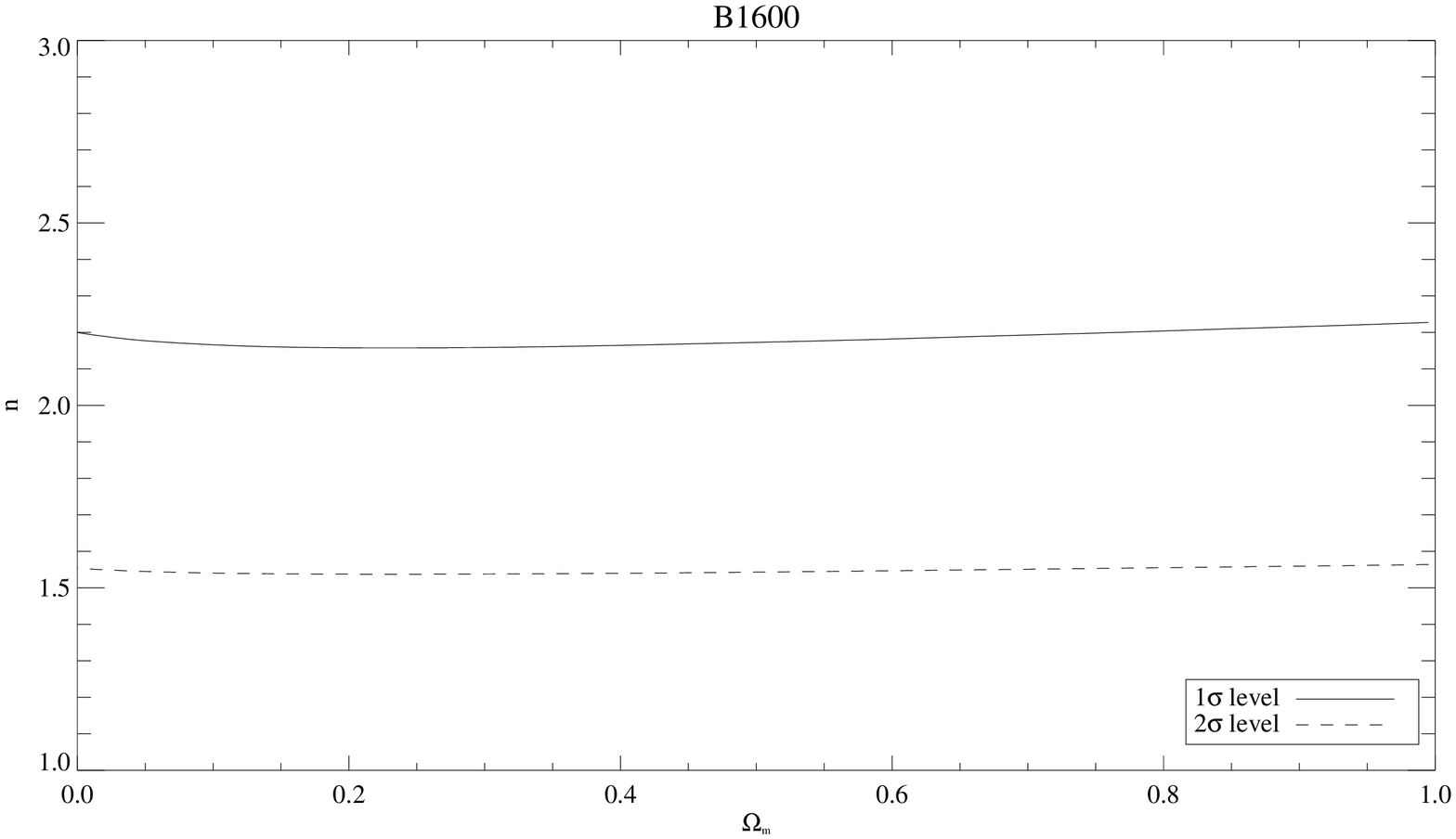}
\caption{\label{omn_b1600} Marginalized $1$ and $2\sigma$ contours in the $\Omega_m-n$ plane without WMAP prior for B1600-434.}
\includegraphics[width=8cm]{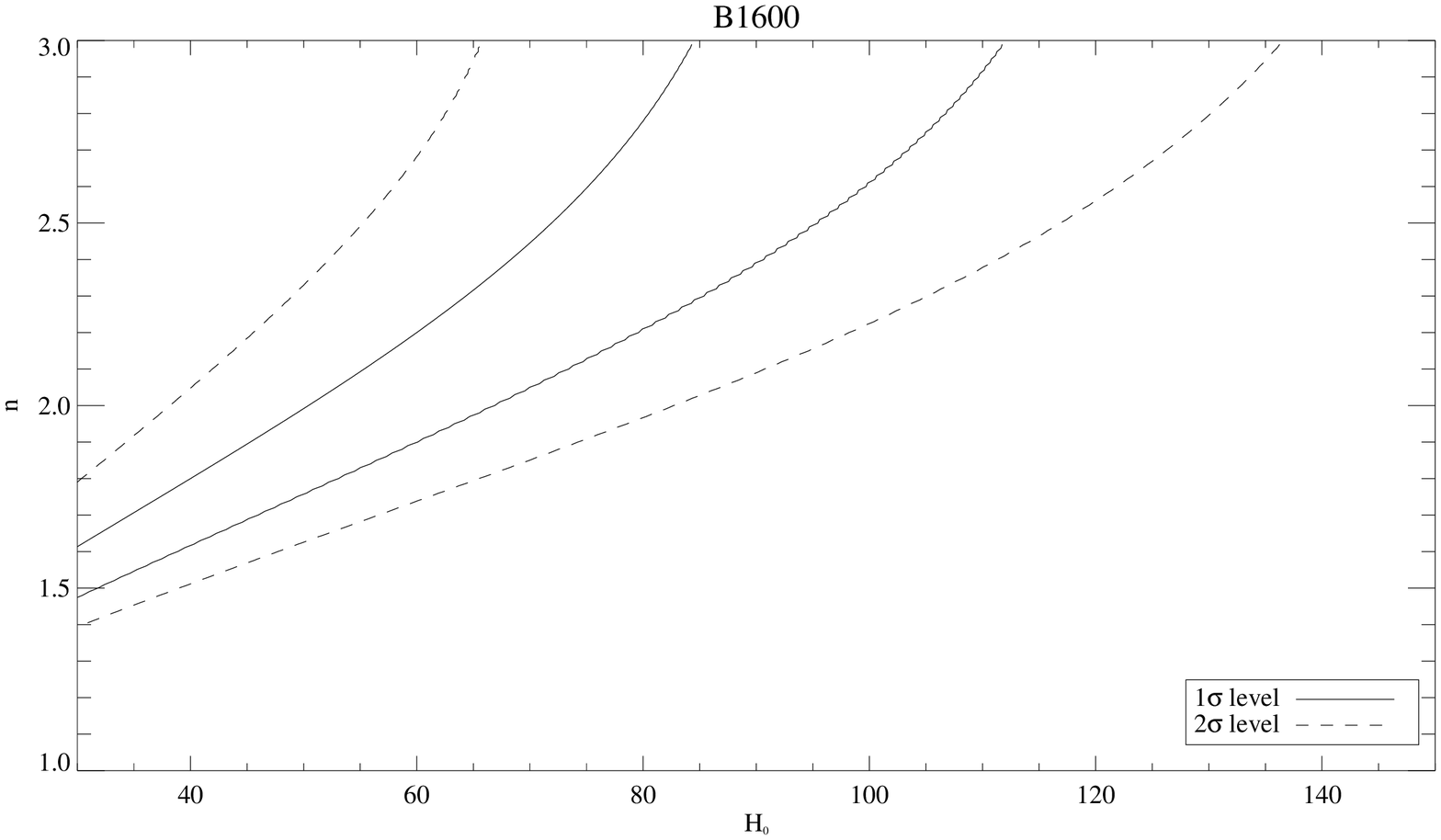}
\caption{\label{h0n_b1600}As in Fig. \ref{omn_b1600} for the $H_0-n$ plane without WMAP prior.}
\end{figure}

Using the image positions and time-delays of B1600+434, SBS~1520+530 and SDSS~J1650+4251, we perform 
a likelihood analysis to 
infer constraints on $H_0$ after marginalizing over the angular errors of the image locations, the slope of 
the power-law model of each lens $n_i$ and the matter density $\Omega_m$.  
We assume flat priors on $1<n<3$ and $30<H_0\,[{\rm km\,s^{-1}Mpc^{-1}}]<150$.  
For each lens we compute the likelihood function $\mathcal{L}(n_i,\Omega_m,H_0)$ 
as given by Eq. (\ref{likeerr}). We illustrate the parameter degeneracies for B1600+434,
in Fig. \ref{omn_b1600} and Fig. \ref{h0n_b1600} we plot the $1$ and $2\sigma$
marginalized contours in the planes $\Omega_m-n$ and $H_0-n$ respectively. We can see that 
there is no correlation between $n$ and $\Omega_m$ for such dataset, while a more pronounced
degeneracy is present between $n$ and $H_0$, as was already pointed in \cite{Suyu}. Introducing a prior on $\Omega_m$ does not change the likelihood contour significantly, as there is little degeneracy between $\Omega_m$ and either $n$ or $H_0$.

\begin{figure}
\includegraphics[width=8cm]{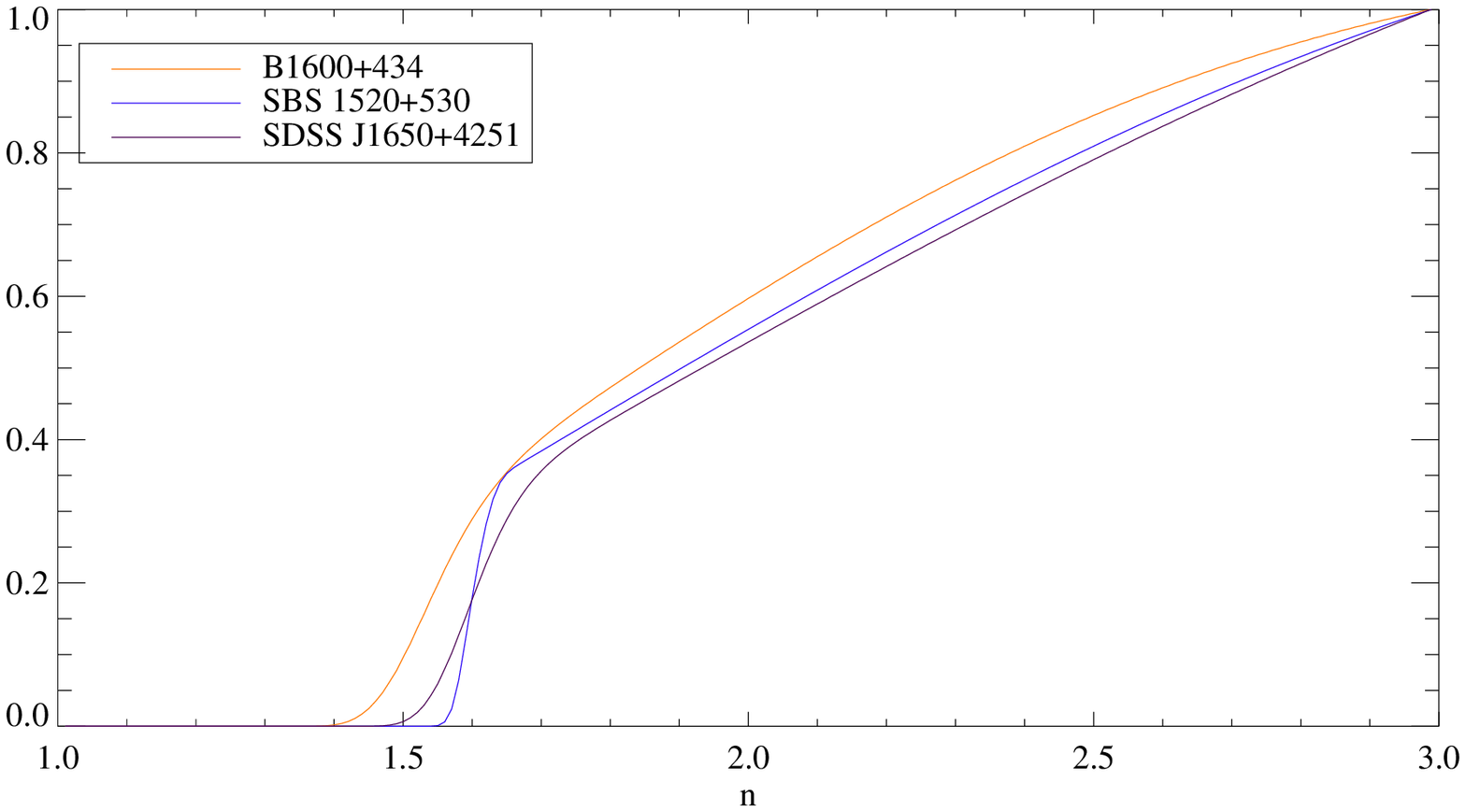}
\caption{\label{n_slope} $\mathcal{L}(n\mid\Omega_m,H_0)/\mathcal{L}_\mathrm{max}$  for each of the selected lenses.}
\includegraphics[width=8cm]{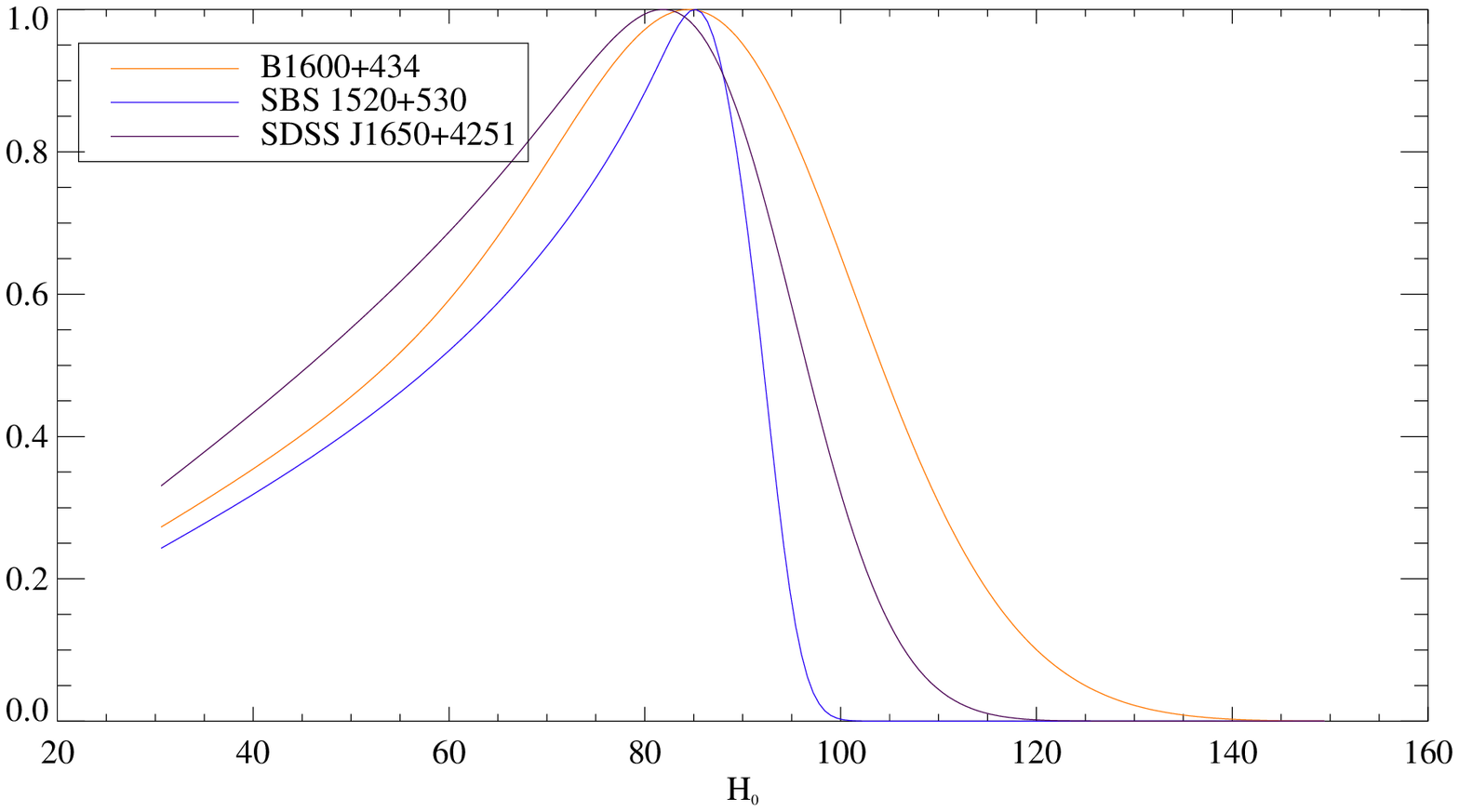}
\caption{\label{h0_lens}$\mathcal{L}(H_0\mid\Omega_m,n)/\mathcal{L}_\mathrm{max}$ for each of the selected lenses.}
\end{figure}

In Fig. \ref{n_slope} we plot the 1-dimensional marginalized likelihoods   
in the prior interval of $n$ for each of the lenses obtained assuming 
a Gaussian prior on $\Omega_m=0.266\pm0.029$ consistent with WMAP7-yrs results \citep{WMAP7}. 
We can see that data can only provide a lower limit on the value of the slope of the lens potential, 
$n\gtrsim 1.5$. Similarly, in Fig. \ref{h0_lens} we plot the marginalized likelihoods of $H_0$ for each lens. 
We may notice that the likelihoods peak around nearly the same value of $H_0$ and 
have a large dispersion around the maximum value.
 
\begin{figure}
\begin{tabular}{c}
\includegraphics[width=7.5cm]{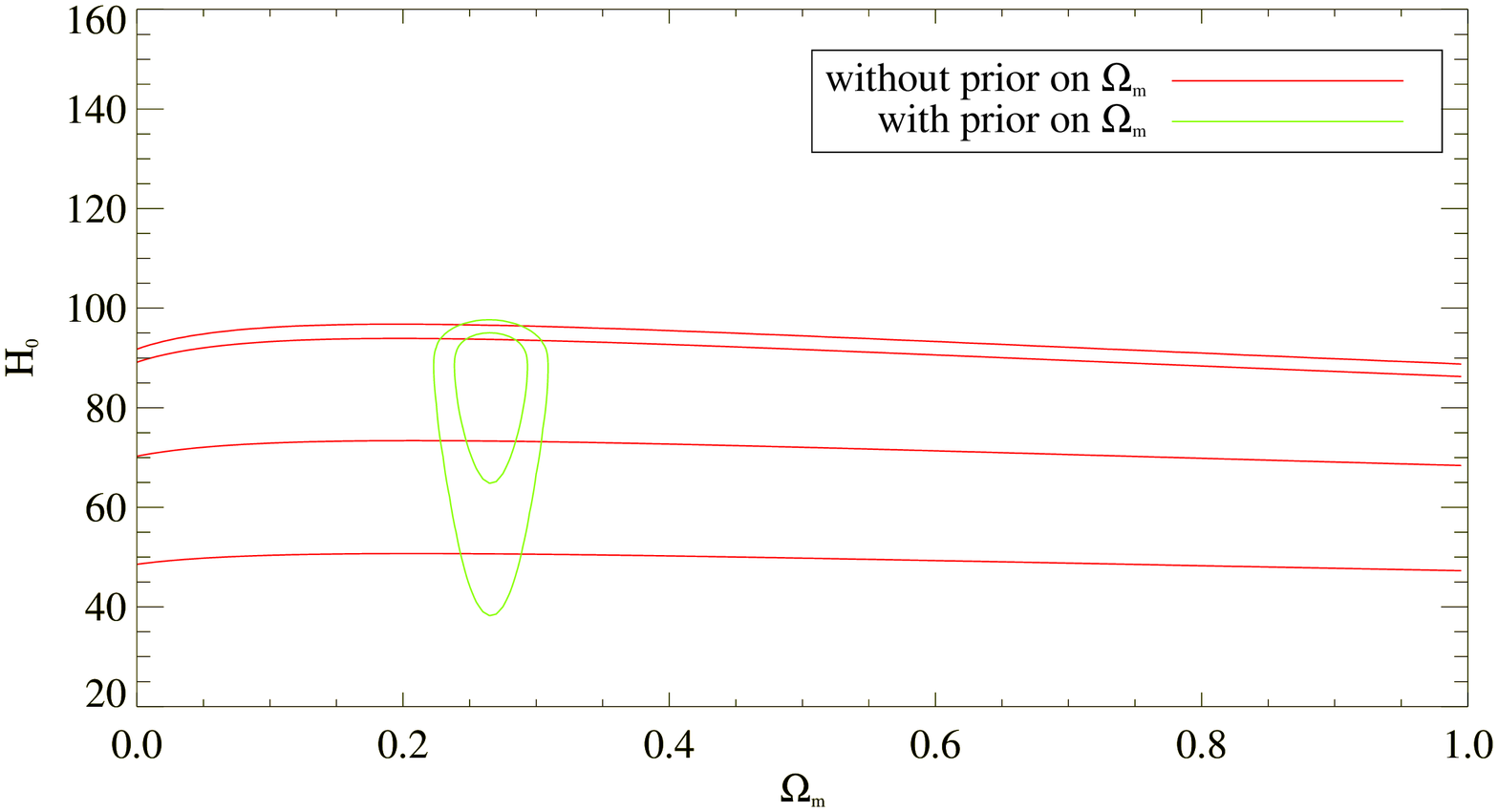}\\
\includegraphics[width=7.5cm]{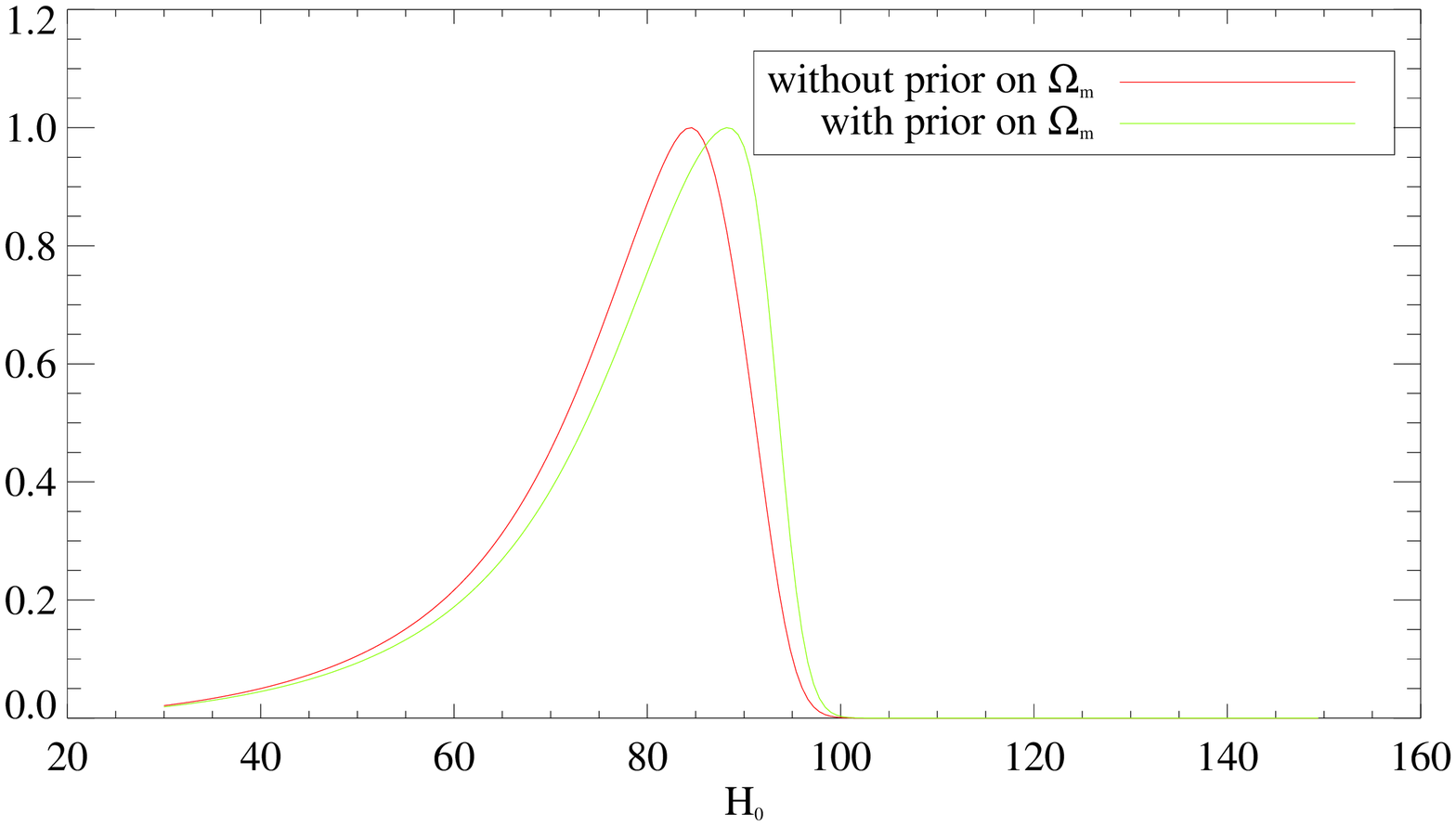}
\end{tabular}
\caption{\label{combinedlike} Top panel: $1$ and $2\sigma$ contours $\Omega_m-H_0$ plane from the combined analysis of the selected lens sample. 
Bottom panel: $\mathcal{L}(H_0)/\mathcal{L}_\mathrm{max}$ for the same sample.}
\end{figure}

The combined constraints on $H_0$ are inferred by computing the total likelihood sample:
\begin{equation}
\mathcal{L}(\Omega_m, H_0)= \prod_i\int \mathcal{L}(n_i,\Omega_m,H_0)\,P(n_i)\,dn_i,
\end{equation}
where the sum is over the likelihood of each lens weighted by $P(n_i)$ the uniform prior in the interval $1<n_i<3$. Hence, we do not assume a hard prior on $n$ at the best-fit value of each lens,
rather we impose the model and propagate its parameter uncertainties by marginalising the likelihood over them.
In Fig. \ref{combinedlike} we plot the $1$ and $2\sigma$ contours in the $\Omega_m-H_0$ plane (left panel) 
and the 1-dimensional marginalized likelihood for $H_0$ (right panel) with and without WMAP7-yrs prior. 
As we can see, time-delays are mostly insensitive to $\Omega_m$. 
The average and standard deviation values are $H_0=76^{+ 15}_{- 5} \, \mathrm{km\,s^{-1}Mpc^{-1}}$ and
$H_0=78^{+ 15}_{- 5}  \, \mathrm{km\,s^{-1}Mpc^{-1}}$ without and with WMAP7-yr prior respectively. Notice
that the marginalized likelihood is non-Gaussian and strongly skewed towards small values of $H_0$. Thus, the average values differ the maximum ones which
are at $H_0 = 85$~km~s$^{-1}$~Mpc$^{-1}$ and $H_0 = 88$~km~s$^{-1}$~Mpc$^{-1}$ without and with WMAP7-yr prior respectively.

It is worth highlighting that, while on the observational sample, about a quarter (three out of twelve) of the lenses are selected through our analysis, in the simulated sample this ratio is higher than a third (111 out of 280), which seems to indicate that observational samples are more contaminated than our simulated sample, leading to possibly higher bias if the whole sample is used to measure $H_0$.

\section{Conclusions}\label{conclusions}
In this study we have used Bayesian model selection techniques to determine the lens mass model which given the data
has the highest probability to describe a strong gravitational lens system. Rather than modeling a lens in all its complexity, our goal is to focus on selecting a model whose 
parameters can significantly influence the time-delay such as to infer unbiased cosmological constraints when averaging individual mass model parameter uncertainties on a homogeneous lens
sample. To this end we have focused on double lenses and used Bayes factor statistics to select a sample of lenses that can be described by a power-law model without external shear.
We have tested this approach on a simulated sample of 500 lenses. The likelihood analysis of the selected subsample assuming no external shear recovers the 
fiducial value of the Hubble constant with $3\sigma$ statistical uncertainty, thus a systematic selection bias is no greater than $\sim 5\%$. On the other hand, since the more complex
model including external shear parameter is underconstrained using double lenses, the likelihood analysis result in strongly biased results due to the marginalized effects
of large lens model parameter degeneracies. Therefore, a given set of data should not be brute force analysed assuming the model capable of accounting for the maximal complexity. 
Rather, performing a Bayes factor analysis may provide a more effective guide to data model selection. 

The application to a sample of observed lenses indicates that
out of the initial dataset, nine have Bayes factors favoring a simple power-law lens model, though six of them lies in the  ``inconclusive'' interval and do not pass our
conservative cut. These are B0218+357, FBQ 0951+2635, HE 2149-2745, PKS 1830-211, Q0142-100, and SBS 0909+532. It is possible that more accurate time-delay measurements
will update their Bayes factor value and give us a better knowledge of the appropriate lens model description.
We have performed the same analysis including information from lens flux-ratio measurements. In such a case, the Bayes factors indicate that external shear must be
taken into account for most of the lenses. However, several source of astrophysical systematic errors may affect the flux-ratios such as dust extinction and microlensing events that do not
reflect the lensing magnification caused by the lens mass distribution responsible for the images and time-delays. Indeed, uncontaminated flux-ratios will provide interesting additional 
constraint on the lens model. Nevertheless, it is worth reminding that time-delays are less sensitive to substructure than flux-ratios and require a less complex modelling.  

Here, we have restricted the analysis to double lens systems. Such systems have fewer constraints than quads, nevertheless the advantage is that they can be described in terms of simpler 
models with fewer parameters. This is not the case for quad lenses in which the quadrupole structure induces an internal shear that needs to be taken into account to correctly describe the lens time-delay.
The lens reconstruction technique described in \cite{Alard2007,Alard2008} and adapted in \cite{Habara2011} could be particularly useful in this case especially in combination with the Bayesian selection
approach described here to determine the order of the perturbative lens reconstruction. Though, such technique may present difficulties since in most cases we do not see arcs but only point-like images.

The number of known lensed quasars which are potentially good targets for time-delay measurements is currently of order a hundred. In the future such dataset will increase thanks to numerous survey
programs. The use of the Bayesian model selection we have discussed can provide large homogeneous subsample of lenses that are suitable for unbiased cosmological parameter inference. 
In particular, along the line of \citet{Parkinson2010}, using Markov Monte Carlo Chain techniques one could perform a Bayesian model averaging of cosmological parameters over homogeneous subsamples of lenses each described by a class of lens models.

\section*{acknowledgements}
We thank A.R. Liddle, D. Sluse, J. Wambsganss, C.R. Keeton and S.H. Suyu for their valuable comments and the anonymous referee who has helped us to improve and enrich the content of this paper. 
I. Balm\`es is supported by a scolarship of the ``Minist\`ere de l'\'Education Nationale, de la Recherche et de la Technologie'' (MENRT). The research leading to these results has received funding from the European Research Council under the European Community's Seventh Framework Programme (FP7/2007-2013 Grant Agreement no. 279954).

\appendix
\section{Model with External Shear}\label{appendixA}
In this Appendix we present a derivation of the time-delay and magnification equations respectively in the case of a 
power-law potential with external shear. The latter reads as 
\begin{equation}
\psi(\btheta) = \frac{b^2}{3-n}\left(\frac{\theta}{b} \right)^{3-n} - \gamma \frac{\theta^2}{2} \cos 2 (\phi -\phi_\gamma).
\end{equation}
We can relate one of the model parameters to the remaining ones using the lens equation, since both images result of the same source at $\bbeta$.
This gives us the following equation:
\begin{equation}
|\bbeta(\btheta_\rmn{A})|^2=|\bbeta(\btheta_\rmn{B})|^2
\end{equation}
which can be rewritten, using the lens equation $\bbeta = \btheta - \bigtriangledown \psi(\btheta)$, as a second order equation in $X=b^{n-1}$:
\begin{equation}
UX^2+2VX+W=0
\end{equation}
with
\begin{eqnarray}
U &=& \theta_\rmn{A}^{2(2-n)}-\theta_\rmn{B}^{2(2-n)} \\
V &=& \theta_\rmn{B}^{3-n}(1+\gamma C_\rmn{B}) - \theta_\rmn{A}^{3-n}(1+\gamma C_\rmn{A}) \\
W &=& \theta_\rmn{A}^2 (1 + 2 \gamma C_\rmn{A} + \gamma^2) - \theta_\rmn{B}^2 (1 + 2 \gamma C_\rmn{B} + \gamma^2)
\end{eqnarray}
where $C_\rmn{A,B}=\cos 2 (\phi_\rmn{A,B} -\phi_\gamma)$.
Assuming that $\theta_\rmn{A} \geq \theta_\rmn{B}$ and $ n > 1$, and using the usual notation $\Delta = V^2 - U W$ we have the following solutions:
for $\Delta > 0$
\begin{equation}
b^{n-1}=\frac{-V + \sqrt{\Delta}}{U},
\end{equation}
the solution $b^{n-1}=\frac{-V - \sqrt{\Delta}}{U}$ is eliminated since it does not reduce to the right value for $\gamma=0$ (note however that it would be the correct solution if $n<1$). Note also that for $n=1$, there is no parameter $b$ and this equation has no meaning;
for $\Delta = 0$, the solution is simply
\begin{equation}
b^{n-1}=\frac{-W}{2V},
\end{equation}
while for $\Delta < 0$, there are no real solutions and the lens cannot be described for this combination of $n$, $\gamma$ and $\phi_\gamma$.
Having reduced the number of lens model parameters the time delay between images A and B can be computed using the value of $t_\rmn{A}$ and $t_\rmn{B}$:
\begin{equation}
t_\rmn{i}=(1+z_\rmn{l}) \frac{D_\rmn{l} D_\rmn{s}}{D_\rmn{ls}} \left[\frac{1}{2}\mid\bigtriangledown \psi\mid^2 + \psi(\btheta_\rmn{i}) \right],
\end{equation}
where we have used the lens equation $\bbeta - \btheta_\rmn{i}= - \bigtriangledown \psi(\btheta_\rmn{i})$.
In the case of the magnification we have for each image:
\begin{eqnarray}
\mu_\rmn{i}^{-1}=\left[1-\left(\frac{\theta_\rmn{i}}{b}\right)^{1-n}\right] \left[1- (2-n)\left(\frac{\theta_\rmn{i}}{b}\right)^{1-n} \right] \\ + (1-n) \gamma C_\rmn{i}\left(\frac{\theta_\rmn{i}}{b}\right)^{1-n} - \gamma^2,
\end{eqnarray}
and the flux ratio is simply given by $F_\rmn{AB}=\mu_\rmn{A}/\mu_\rmn{B}$.

\section{Lens Data Astrometry}\label{appastrometry}
In table~\ref{astrometry} we report the astrometric data of the positions of images (A,B) and lens (L) for all lenses used in the lens data analysis. 
The references to the data are listed in table~\ref{data}.

\begin{table}
\caption{\label{astrometry} Astrometry for all lenses used in the data analysis}
\centering
\begin{tabular}{cccc}
\hline
lens & object & $\Delta\alpha$ & $\Delta\delta$ \\ \hline

B0218+357 & L & $\equiv 0$ & $\equiv 0$ \\
& A & $-0.250 \pm 0.005$ & $-0.125 \pm 0.007$ \\
& B & $0.057 \pm 0.006$ & $0.001 \pm 0.008$ \\

B1600+434 & L & $\equiv 0 \pm 0.05$ & $\equiv 0 \pm 0.05$ \\
& A & $-0.33 \pm 0.01$ & $1.09 \pm 0.01$ \\
& B & $0.07 \pm 0.01$ & $-0.24 \pm 0.01$ \\

FBQ 0951+2635 & G & $0.760 \pm 0.003$ & $-0.455 \pm 0.003$ \\
& A & $\equiv 0$ & $\equiv 0$ \\
& B & $0.900 \pm 0.003$ & $-0.635 \pm 0.003$ \\

HE 1104-1805 & G & $0.974 \pm 0.003$ & $-0.510 \pm 0.004$ \\
& A & $\equiv 0$ & $\equiv 0$ \\
& B & $2.901 \pm 0.003$ & $-1.332 \pm 0.003$ \\

HE 2149-2745 & G & $0.714 \pm 0.007$ & $1.150 \pm 0.005$ \\
& A & $\equiv 0$ & $\equiv 0$ \\
& B & $0.890 \pm 0.003$ & $1.446 \pm 0.003$ \\

PKS 1830-211 & G & $0.498 \pm 0.004$ & $-0.456 \pm 0.004$ \\
& A & $\equiv 0$ & $\equiv 0$ \\
& B & $0.649 \pm 0.001$ & $-0.724 \pm 0.001$ \\

Q0142-100 & G & $1.764 \pm 0.003$ & $-0.574 \pm 0.003$ \\
& A & $\equiv 0$ & $\equiv 0$ \\
& B & $2.145 \pm 0.003$ & $-0.613 \pm 0.003$ \\ 

Q0957+561 & G & $1.406 \pm 0.006$ & $-5.027 \pm 0.005$ \\
& A & $\equiv 0$ & $\equiv 0$ \\
& B & $1.229 \pm 0.005$ & $-6.048 \pm 0.004$ \\

SBS 0909+532 & G & $0.415 \pm 0.125$ & $-0.004 \pm 0.081$ \\
& A & $\equiv 0$ & $\equiv 0$ \\
& B & $0.987 \pm 0.003$ & $-0.498 \pm 0.003$ \\

SBS 1520+530  & G & $1.141 \pm 0.003$ & $-0.395 \pm 0.003$ \\
& A & $\equiv 0$ & $\equiv 0$ \\
& B & $1.429 \pm 0.003$ & $-0.652 \pm 0.003$ \\

SDSS J1206+4332 & G & $-0.664 \pm 0.137$ & $1.748 \pm 0.028$ \\
& A & $\equiv 0 \pm 0.011$ & $\equiv 0 \pm 0.010$ \\
& B & $-0.098 \pm 0.006$ & $2.894 \pm 0.009$ \\

SDSS J1650+4251  & G & $0.017 \pm 0.032$ & $-0.872 \pm 0.026$ \\
& A & $\equiv 0$ & $\equiv 0$ \\
& B & $0.223 \pm 0.002$ & $-1.163 \pm 0.001$ \\ \hline

\end{tabular}
\end{table}

\end{document}